\documentclass[aps]{revtex4}

\usepackage{graphicx}
\usepackage{natbib}

\usepackage{graphics}
\usepackage{epsfig}
\usepackage{bm}
\usepackage{amsmath,amssymb}
\usepackage{stmaryrd}

\def\wt#1{\widetilde{#1}}
\def\vb#1{\mbox{\boldmath$#1$}}
\def\pd#1#2{\frac{\partial #1}{\partial #2}}
\def\fd#1#2{\frac{\delta #1}{\delta #2}}
\def\wh#1{\widehat{#1}}
\def\bdot{\,\vb{\cdot}\,}
\def\btimes{\,\vb{\times}\,}

\def\bhat{\wh{{\sf b}}}
\def\cal#1{\mathcal{#1}}

\def\exd{{\sf d}}
\def\bhat{\wh{{\sf b}}}

\newcommand{\bc}{\begin{center}}
\newcommand{\ec}{\end{center}}
\newcommand{\bt}{\begin{tabbing}}
\newcommand{\et}{\end{tabbing}}
\newcommand{\be}{\begin{equation}}
\newcommand{\ee}{\end{equation}}
\newcommand{\ba}{\begin{eqnarray}}
\newcommand{\ea}{\end{eqnarray}}

\begin{document}

\title{Symplectic gyrokinetic Vlasov-Maxwell Theory}

\author{A. J. Brizard}
\affiliation{Department of Physics, Saint Michael's College, Colchester, VT 05439, USA}

\begin{abstract} 
A new representation of electromagnetic gyrokinetic Vlasov-Maxwell theory is presented in which the gyrocenter equations of motion are expressed solely in terms of the perturbed electric and magnetic fields. In this representation, the gyrocenter symplectic (Poisson-bracket) structure and the gyrocenter Jacobian contain electric and magnetic perturbation terms associated with the standard first-order gyrocenter polarization and magnetization terms that traditionally appear in the gyrokinetic Maxwell equations. In addition, the gyrocenter polarization drift (which includes perturbed  magnetic-field corrections) now appears explicitly in the gyrocenter velocity. The symplectic gyrokinetic Vlasov-Maxwell equations are self-consistently derived from a constrained Eulerian variational principle, which yields exact energy-momentum conservation laws (through the Noether method) that are verified explicitly. An exact toroidal canonical angular momentum conservation law is also derived explicitly under the assumption of an axisymmetric background magnetic field.
\end{abstract}

\date{\today}


\maketitle

\section{Introduction}

Gyrokinetics was originally introduced by Taylor \cite{Taylor_1967} as a procedure by which the adiabatic invariance of the (guiding-center) magnetic moment, which is destroyed in the presence of low-frequency, short-wavelength electrostatic perturbations, could be restored by introducing an additional asymptotic expansion in powers of the perturbation-field amplitude. Since the adiabatic invariance of the magnetic moment is the foundation of magnetic plasma confinement \citep{Cary_Brizard_2009}, Taylor's work was a crucial first step in investigating the stability of magnetized plasma equilibria in the presence of the low-frequency, short-wavelength perturbations assumed to be responsible for anomalous plasma transport. 

The subsequent development of linear \citep{Catto_1978,Catto_1981} and nonlinear \citep{Frieman_Chen_1982} gyrokinetic equations initiated an extensive research program drawing on their analytical and numerical properties \citep{Brizard_Hahm_2007,Garbet_2010,Krommes_2012}. In these gyrokinetic numerical applications, the exact conservation of gyrokinetic energy served as a guide toward the development energy-conserving numerical algorithms used in studying saturated turbulent transport in magnetized plasmas \citep{Brizard_Hahm_2007}. We note that, while the gyrokinetic energy conservation law was initially derived directly from the gyrokinetic Vlasov-Poisson equations \citep{Dubin_1983} and gyrokinetic Vlasov-Maxwell equations \citep{HLB_1988,Brizard_1989}, the existence of variational principles \citep{Sugama_2000,Brizard_PRL_2000,Brizard_PoP_2000,Brizard_2010} for these gyrokinetic equations has led to a more systematic derivation of the gyrokinetic energy conservation law.

One of the hallmarks of Hamiltonian gyrokinetic Vlasov-Maxwell theory \citep{Brizard_Hahm_2007} involves the polarization and magnetization effects appearing in the gyrokinetic Maxwell equations. In standard Hamiltonian gyrokinetic Vlasov-Maxwell theory, the first-order gyrocenter polarization and magnetization are derived by variations of the second-order gyrocenter Hamiltonian with respect to the first-order electric and magnetic fields, respectively. In a self-consistent Hamiltonian theory, the second-order gyrocenter Hamiltonian must also appear in the full gyrokinetic Vlasov equation in which second-order gyrocenter drifts are retained in order to satisfy exact energy and momentum conservation laws.

The inclusion of a second-order gyrocenter Hamiltonian, however, can be cumbersome for practical applications in gyrokinetic particle simulations and alternate gyrokinetic models (such as the $\delta f$ representation) or truncated models \citep{Mandell_2020} are often preferred; see the recent reviews by Garbet {\it et al.} \cite{Garbet_2010} and Krommes \cite{Krommes_2012}. In addition, the standard gyrokinetic equations are expressed in terms of the perturbed electromagnetic potentials $(\Phi_{1}, {\bf A}_{1})$ which, because of their gauge dependence, offers limited applications when only perturbed electromagnetic fields $({\bf E}_{1},{\bf B}_{1})$ are involved. 

The purpose of the present work is to offer a new gauge-invariant gyrokinetic model whose equations are expressed solely in terms of the perturbed electromagnetic fields $({\bf E}_{1},{\bf B}_{1})$. In addition, the work explores the role played by polarization and magnetization in establishing exact conservation laws for energy-momentum and angular momentum, which are self-consistently derived from a constrained Eulerian variational principle \citep{Brizard_PRL_2000,Brizard_PoP_2000}. Within this context, we also explore the connection between the second-order gyrocenter Hamiltonian and these gyrokinetic conservation laws. Lastly, each gyrokinetic conservation law is proved explicitly in order to motivate and justify the applications of Noether's method in gyrokinetic Vlasov-Maxwell theory.

\section{Polarization and Magnetization in Vlasov-Maxwell Theory}

Over the past decade \citep{Brizard_2008,Brizard_2009,Brizard_Tronci_2016,Brizard_2018,EH_2020}, the variational formulation of generic reduced Vlasov-Maxwell equations has revealed the deep connection between reduced polarization and magnetization effects, on the one hand, and the energy-momentum conservation laws, on the other hand. These results have generalized previous works by Pfirsch \cite{Pfirsch_1984} and Pfirsch \& Morrison \cite{Pfirsch_Morrison_1985} for the case of the guiding-center Vlasov-Maxwell equations. In recent work by Brizard \& Tronci \cite{Brizard_Tronci_2016}, the guiding-center polarization and magnetization have played a crucial role in demonstrating exact energy-momentum conservation properties. In particular, for the first time, the guiding-center stress tensor has been demonstrated to be explicitly symmetric, which then implies the existence of an exact angular-momentum conservation law. We note that guiding-center Vlasov-Maxwell theory is different from drift-kinetic Vlasov-Maxwell theory \citep{Pfirsch_1984,Dimits_1992}, where in the former theory, the electromagnetic fields are not separated into contributions from a time-independent background magnetic field and time-dependent electromagnetic perturbation fields, with both contributions satisfying the guiding-center space-time orderings.

\subsection{Polarization and magnetization in guiding-center theory}

Before deriving our new gyrokinetic Vlasov-Maxwell model, we begin with a review of polarization and magnetization effects in guiding-center Vlasov-Maxwell theory  \citep{Pfirsch_1984,Pfirsch_Morrison_1985,Brizard_Tronci_2016,Cary_Brizard_2009,Brizard_2013,Tronko_Brizard_2015,Brizard_2017a}. For this purpose, we introduce the generic guiding-center Lagrangian expressed in terms of the guiding-center phase-space coordinates $({\bf X},p_{\|},J \equiv (mc/e)\,\mu,\zeta)$:
\begin{eqnarray}
L_{\rm gc} & = & \left( \frac{e}{c}\,{\bf A} + p_{\|}\,\bhat + \vb{\Pi}_{\rm gc}\right)\bdot\dot{\bf X} + J\,\dot{\zeta} \;-\; \left( \frac{p_{\|}^{2}}{2m} \;+\; \mu\,B \;+\; e\,\Phi \;+\; e\,\Psi_{\rm gc}\right) \nonumber \\
 & \equiv & {\bf P}_{\rm gc}\bdot\dot{\bf X} + J\,\dot{\zeta} - H_{\rm gc},
\label{eq:L_gc}
\end{eqnarray}
where the magnetic field is ${\bf B} \equiv \nabla\btimes{\bf A} = B\,\bhat$, while the guiding-center symplectic momentum $\vb{\Pi}_{\rm gc}$ and the guiding-center potential $\Psi_{\rm gc}$ may depend on the time-dependent electromagnetic fields $({\bf E},{\bf B})$.

In guiding-center Vlasov-Maxwell theory \citep{Cary_Brizard_2009}, the guiding-center polarization and magnetization are defined, respectively, as guiding-center momentum-moments of the guiding-center Vlasov distribution $F({\bf X},p_{\|},J,t)$ and derivatives of the guiding-center Lagrangian \eqref{eq:L_gc} with respect to $({\bf E}, {\bf B})$ at fixed $({\bf X},\dot{\bf X})$:
\begin{eqnarray}
\mathbb{P}_{\rm gc}({\bf X},t) & \equiv & \int_{\bf P}{\cal J}_{\rm gc}\,F\;\pd{L_{\rm gc}}{{\bf E}} \;=\; \int_{\bf P}{\cal J}_{\rm gc}\,F\;\left(\pd{\vb{\Pi}_{\rm gc}}{{\bf E}}\bdot\dot{\bf X} \;-\; e\;\pd{\Psi_{\rm gc}}{{\bf E}} \right),
\label{eq:Polgc_def} \\
\mathbb{M}_{\rm gc}({\bf X},t) & \equiv & \int_{\bf P}{\cal J}_{\rm gc}\,F\;\pd{L_{\rm gc}}{{\bf B}} \;=\; \int_{\bf P}{\cal J}_{\rm gc}\,F\;\left(\pd{{\bf P}_{\rm gc}}{{\bf B}}\bdot\dot{\bf X} \;-\; \pd{H_{\rm gc}}{{\bf B}} \right) \nonumber \\
 & = & \int_{\bf P}{\cal J}_{\rm gc}\,F\;\left[ \left(\frac{e\bhat}{\Omega}\btimes\dot{\bf X}\right)\btimes\frac{p_{\|}\bhat}{mc} + \pd{\vb{\Pi}_{\rm gc}}{{\bf B}}\bdot\dot{\bf X}  - \left(\mu\,\bhat  + e\;\pd{\Psi_{\rm gc}}{{\bf B}} \right) \right],
\label{eq:Maggc_def} 
\end{eqnarray}
where ${\cal J}_{\rm gc}$ denotes the guiding-center Jacobian. We note that, although the Jacobian may also depend on the electromagnetic fields $({\bf E}, {\bf B})$, it is not involved in the definition of polarization and magnetization. In Eqs.~\eqref{eq:Polgc_def}-\eqref{eq:Maggc_def}, we note that polarization and magnetization effects arise from two separate contributions: from field dependence of the canonical (symplectic) momentum ${\bf P}_{\rm gc}$ and field dependence of the Hamiltonian $H_{\rm gc}$. We also note that the guiding-center magnetization \eqref{eq:Maggc_def} has contributions that are independent of the momentum-energy functions $(\vb{\Pi}_{\rm gc}, \Psi_{\rm gc})$. 

In a recent review on Hamiltonian guiding-center theory by Cary \& Brizard \cite{Cary_Brizard_2009}, two different guiding-center Lagrangians were presented:
\begin{equation}
L_{\rm gc} = \left\{ \begin{array}{l}
\left( e{\bf A}/c + p_{\|}\,\bhat\right)\bdot\dot{\bf X} + J\,\dot{\zeta} - \left(p_{\|}^{2}/2m + \mu B + e \Phi - \vb{\pi}_{\rm gc}\bdot{\bf E} - m|{\bf u}_{E}|^{2}/2\right) \\
 \\
\left( e{\bf A}/c  + p_{\|}\,\bhat + m{\bf u}_{E}\right)\bdot\dot{\bf X} + J\,\dot{\zeta} - \left(p_{\|}^{2}/2m + \mu B + e \Phi + m|{\bf u}_{E}|^{2}/2\right) 
\end{array} \right.
\label{eq:Lag_gc}
\end{equation}
where the top Lagrangian includes the $E\times B$ velocity (${\bf u}_{E} = {\bf E}\btimes c\bhat/B$) in the Hamiltonian only (i.e., $\vb{\Pi}_{\rm gc} = 0$) and $e\Psi_{\rm gc} = - \vb{\pi}_{\rm gc}\bdot{\bf E} - m|{\bf u}_{E}|^{2}/2$ includes the guiding-center electric-dipole moment $\vb{\pi}_{\rm gc}$ generated by magnetic-field gradient and curvature effects \citep{Tronko_Brizard_2015,Brizard_Tronci_2016,Brizard_2017a}. The bottom Lagrangian, on the other hand, includes the $E\times B$ velocity in both the symplectic and Hamiltonian parts $\vb{\Pi}_{\rm gc} = m{\bf u}_{E}$ and $e\Psi_{\rm gc} = m|{\bf u}_{E}|^{2}/2$ \citep{Pfirsch_Morrison_1985}. 

The Euler-Lagrange equations associated with the generic guiding-center Lagrangian \eqref{eq:L_gc} are expressed as
$\dot{p}_{\|}\bhat - \dot{\bf X}\btimes e{\bf B}^{*}/c = e\,{\bf E}^{*}$ and $\bhat\bdot\dot{\bf X} = p_{\|}/m$, which yield the guiding-center equations of motion \citep{Littlejohn_1983,Cary_Brizard_2009}
\begin{equation}
\dot{\bf X} \;=\; \frac{p_{\|}}{m}\;\frac{{\bf B}^{*}}{B_{\|}^{*}} \;+\; {\bf E}^{*}\btimes\frac{c\bhat}{B_{\|}^{*}} \;\;{\rm and}\;\; \dot{p}_{\|} \;=\; e\,{\bf E}^{*}\bdot\frac{{\bf B}^{*}}{B_{\|}^{*}},
\end{equation}
where 
\begin{equation}
\left. \begin{array}{rcl}
e\,{\bf E}^{*} & = & e\,{\bf E} \;-\; \left( \mu\,\nabla B + e\,\nabla\Psi_{\rm gc}\right) \;-\; \left(p_{\|}\,\partial\bhat/\partial t + \partial\vb{\Pi}_{\rm gc}/\partial t\right) \\
e{\bf B}^{*}/c & = & e{\bf B}/c \;+\; \nabla\btimes\left(p_{\|}\,\bhat + \vb{\Pi}_{\rm gc}\right) \\
 eB_{\|}^{*}/c & = & \bhat\bdot e{\bf B}^{*}/c \;=\; eB/c + \bhat\bdot\nabla\btimes\left(p_{\|}\,\bhat + \vb{\Pi}_{\rm gc}\right)
\end{array} \right\}.
\end{equation}
We note that the symplectic (bottom) Lagrangian \eqref{eq:Lag_gc} is constructed specifically so that the guiding-center velocity includes the polarization drift velocity $\partial\vb{\Pi}_{\rm gc}/\partial t = m\,\partial{\bf u}_{E}/\partial t$ \citep{Pfirsch_1984,Pfirsch_Morrison_1985,PCR_2004}. 

From the guiding-center Lagrangians \eqref{eq:Lag_gc}, we first calculate the guiding-center polarization kernels in Eq.~\eqref{eq:Polgc_def}:
\begin{equation}
\pd{\vb{\Pi}_{\rm gc}}{{\bf E}}\bdot\dot{\bf X} \;-\; e\;\pd{\Psi_{\rm gc}}{{\bf E}}  \;=\; \left\{ \begin{array}{l}
(e\bhat/\Omega)\btimes{\bf u}_{E} \;+\; \vb{\pi}_{\rm gc} \;=\; (e\bhat/\Omega)\btimes\dot{\bf X} \\
\\
(e\bhat/\Omega)\btimes\left(\dot{\bf X} \;-\; {\bf u}_{E}\right)
\end{array} \right.
\end{equation}
where the Hamiltonian (top) result combines the full contribution from the $E\times B$ and magnetic drift velocities, while the symplectic (bottom) result
\begin{eqnarray} 
\frac{e\bhat}{\Omega}\btimes\left(\dot{\bf X} \;-\; {\bf u}_{E}\right) &=& \frac{e\bhat}{\Omega}\btimes\left[ \frac{c\bhat}{eB_{\|}^{*}}\btimes\left( \mu\,\nabla B \;+\; p_{\|}\,\frac{d\bhat}{dt} \;+\; m\,\frac{d{\bf u}_{E}}{dt} \right) \right] \nonumber \\
 &\equiv& \vb{\pi}_{\rm gc} \;-\; \frac{e\bhat}{\Omega}\btimes\left(\frac{d{\bf u}_{E}}{dt}\btimes\frac{\bhat}{\Omega_{\|}^{*}}\right)
 \end{eqnarray}
replaces the standard electric-dipole-moment $e\vb{\rho}_{E}\;(= e\bhat\btimes{\bf u}_{E}/\Omega)$  polarization term with the polarization-drift contribution $(\Omega_{\|}^{*})^{-1}d{\bf u}_{E}/dt$, where $d/dt = \partial/\partial t + (p_{\|}\bhat + \vb{\Pi}_{\rm gc})/m\bdot\nabla$ and $\Omega_{\|}^{*} = eB_{\|}^{*}/mc$. In fact, in the symplectic case, the electric-dipole moment $\vb{\rho}_{E}$ has been transferred to the guiding-center Jacobian: 
\begin{equation}
{\cal J}_{\rm gc} \;\equiv\; {\cal J}_{0} \;-\; \nabla\bdot\left( \vb{\rho}_{E}\frac{}{} {\cal J}_{0} \right) \;+\; (p_{\|}\bhat + m{\bf u}_{E})\bdot\nabla\btimes\bhat,
\label{eq:gcJac_E}
\end{equation}
where ${\cal J}_{0} = (e/c)\,B$ is the local particle Jacobian. 

Pfirsch \cite{Pfirsch_1984} and Pfirsch \& Morrison \cite{Pfirsch_Morrison_1985} have shown that, for the case of the symplectic (bottom) guiding-center Lagrangian, the calculation of the guiding-center magnetization kernel in Eq.~\eqref{eq:Maggc_def} yielded more complex expressions. Here, we simply note that, in the absence of an electric field \citep{Brizard_Tronci_2016}, the guiding-center magnetization \eqref{eq:Maggc_def} is generated by two separate contributions: the intrinsic magnetization, generated by the magnetic-dipole moment $-\mu\bhat$, and the moving electric-dipole contribution generated by $(e\bhat/\Omega\btimes\dot{\bf X})\btimes p_{\|}\bhat/mc$, which involves the guiding-center electric-dipole moment $(e\bhat/\Omega\btimes\dot{\bf X})$ moving at the lowest-order guiding-center velocity $(p_{\|}/m)\bhat$. In the work of Brizard \& Tronci  \cite{Brizard_Tronci_2016}, where $(\vb{\Pi}_{\rm gc} = 0, \Psi_{\rm gc} = 0)$, these guiding-center magnetization contributions played an important role in establishing exact and explicit energy-momentum and angular-momentum conservation laws.

\subsection{Polarization and magnetization effects in gyrokinetic theory}

With the guiding-center polarization and magnetization effects henceforth associated with the time-independent background magnetic field ${\bf B}_{0} = \nabla\btimes{\bf A}_{0} = B_{0}\,\bhat_{0}$, with ${\bf E}_{0} = 0$ used in the simplest gyrokinetic model, we now turn our attention to a general discussion of polarization and magnetization effects in gyrokinetic theory associated with time-dependent electromagnetic-field perturbations whose amplitudes are represented by the ordering parameter $\epsilon$. For this purpose, we introduce the generic gyrocenter Lagrangian expressed in terms of the gyrocenter phase-space coordinates $({\bf X},p_{\|},J,\zeta)$:
\begin{equation}
L_{\rm gy} \;\equiv\; \left( \frac{e}{c}\,{\bf A}_{0}^{*} + p_{\|}\,\bhat_{0} + \vb{\Pi}_{\rm gy}\right)\bdot\dot{\bf X} + J\,\dot{\zeta} \;-\; \left( \frac{p_{\|}^{2}}{2m} \;+\; \mu\,B_{0} \;+\; e\,\Psi_{\rm gy}\right),
\label{eq:Gamma_gy}
\end{equation}
where the unperturbed vector potential ${\bf A}_{0}^{*}$ contains higher-order guiding-center corrections \citep{Brizard_2013,Tronko_Brizard_2015}, while the gyrocenter symplectic momentum $\vb{\Pi}_{\rm gy} = \epsilon\,\vb{\Pi}_{1{\rm gy}} + \cdots$ and the gyrocenter potential $\Psi_{\rm gy} = \epsilon\,\Psi_{1{\rm gy}} + \cdots$ may include first-order electromagnetic potential perturbations $(\Phi_{1},{\bf A}_{1})$ and field perturbations $({\bf E}_{1},{\bf B}_{1})$, which are selected on the basis of specific theoretical or numerical considerations. Here, the gyrocenter phase-space coordinates $({\bf X},p_{\|})$ are used to describe the reduced gyrocenter Hamiltonian dynamics, while the fast gyromotion is represented by the action-angle coordinates $(J,\zeta)$, where the gyroaction $J$ is an invariant of the gyrocenter motion since the gyroangle $\zeta$ is an ignorable coordinate of the reduced gyrocenter Hamiltonian dynamics.

In gyrokinetic theory, the gyrocenter polarization and magnetization are defined, respectively, as gyrocenter phase-space moments of the gyrocenter Vlasov distribution $F({\bf X},p_{\|},J,t)$ and functional variations of the gyrocenter Lagrangian
\eqref{eq:Gamma_gy} with respect to the field perturbations ${\bf E}_{1}({\bf x},t)$ and ${\bf B}_{1}({\bf x},t)$ at a fixed position ${\bf x}$ and time $t$:
\begin{eqnarray}
\mathbb{P}_{\rm gy}({\bf x},t) & \equiv & \int_{\cal Z}{\cal J}_{\rm gy}\,F\;\epsilon^{-1}\fd{L_{\rm gy}}{{\bf E}_{1}({\bf x})} \;=\; \int_{\cal Z}{\cal J}_{\rm gy}\,F\;\epsilon^{-1}\left(\fd{\vb{\Pi}_{\rm gy}}{{\bf E}_{1}({\bf x})}\bdot\dot{\bf X} \;-\; e\;
\fd{\Psi_{\rm gy}}{{\bf E}_{1}({\bf x})} \right),
\label{eq:Polgy_def} \\
\mathbb{M}_{\rm gy}({\bf x},t) & \equiv & \int_{\cal Z}{\cal J}_{\rm gy}\,F\;\epsilon^{-1}\fd{L_{\rm gy}}{{\bf B}_{1}({\bf x})} \;=\; \int_{\cal Z}{\cal J}_{\rm gy}\,F\;\epsilon^{-1}\left(\fd{\vb{\Pi}_{\rm gy}}{{\bf B}_{1}({\bf x})}\bdot\dot{\bf X} \;-\; e\;
\fd{\Psi_{\rm gy}}{{\bf B}_{1}({\bf x})} \right),
\label{eq:Molgy_def} 
\end{eqnarray}
where ${\cal J}_{\rm gy}$ denotes the gyrocenter Jacobian, which may also depend on the perturbation fields $({\bf E}_{1},{\bf B}_{1})$. In Eqs.~\eqref{eq:Polgy_def}-\eqref{eq:Molgy_def}, we note that, like the guiding-center definitions 
\eqref{eq:Polgc_def}- \eqref{eq:Maggc_def}, the gyrokinetic polarization and magnetization effects arise from two separate contributions: from field variations of the symplectic momentum $\delta\vb{\Pi}_{\rm gy}$ and field variations of the gyrocenter potential $\delta\Psi_{\rm gy}$. In standard Hamiltonian gyrokinetic Vlasov-Maxwell theory \citep{Brizard_Hahm_2007}, in which the symplectic momentum $\vb{\Pi}_{\rm gy} \equiv 0$, the gyrocenter polarization and magnetization arise solely from the effective gyrocenter potential $\Psi_{\rm gy}$. 

For our present purposes, we will consider a first-order gyrocenter symplectic momentum $\vb{\Pi}_{1{\rm gy}}$ that may depend on any combinations of the following first-order perturbation terms 
\begin{equation}
\left. \begin{array}{ll}
{\rm (i)} & e\,\langle {\bf A}_{1{\rm gc}}\rangle/c \\
{\rm (ii)} & \langle{\bf E}_{1\bot{\rm gc}}\rangle\btimes(e\bhat_{0}/\Omega) \\
{\rm (iii)} & p_{\|}\,\langle{\bf B}_{1\bot{\rm gc}}\rangle/B_{0}
\end{array} \right\},
\label{eq:scenarios}
\end{equation}
or none at all, in the so-called Hamiltonian representation \citep{Brizard_1989}. Here, the first-order perpendicular electric and magnetic vector fields  ${\bf E}_{1\bot{\rm gc}} = -\,\nabla_{\bot}\Phi_{1{\rm gc}}$ and ${\bf B}_{1\bot{\rm gc}} = \nabla_{\bot}A_{1\|{\rm gc}}\btimes\bhat_{0}$ are expressed in terms of the lowest-order guiding-center and gyrokinetic orderings \citep{Brizard_Hahm_2007}. In addition, a perturbed field ${\sf T}_{\rm gc}^{-1}f_{1} = f_{1{\rm gc}} \equiv f_{1}({\bf X} + \vb{\rho}_{0},t)$ is transformed into a function on guiding-center phase space with the help of the push-forward Lie transform ${\sf T}_{\rm gc}^{-1} \equiv \exp(\vb{\rho}_{0}\bdot\nabla)$, which is expressed in terms of the lowest-order guiding-center transformation \citep{Cary_Brizard_2009} (involving the gyroangle-dependent guiding-center gyroradius $\vb{\rho}_{0}$) and $\langle f_{1{\rm gc}}\rangle({\bf X},J)$ denotes the gyroangle-averaged part of $f_{1{\rm gc}}$. The parallel-symplectic scenario, where only the parallel component $A_{1\|{\rm gc}} = \bhat_{0}\bdot{\bf A}_{1{\rm gc}}$ is kept in scenario (i), was initially considered by Hahm, Lee, \& Brizard \cite{HLB_1988} and was recently discussed by Brizard \cite{Brizard_2017b} within a variational formulation; the polarization-drift scenario (ii) was discussed in Refs.~\cite{Wang_Hahm_2010a,Wang_Hahm_2010b,Leerink_2010,Heikkinen_Nora_2011}; and the combined scenarios (i)-(ii) was discussed by Duthoit {\it et al.} \cite{Duthoit_2014}. The scenario (iii), which involves adding the magnetic flutter velocity to the gyrocenter symplectic structure, has not yet been discussed in the literature until now. In a recent paper, Burby \& Brizard \cite{Burby_Brizard_2019} derived a gauge-invariant gyrokinetic theory by using the local minimal-coupling terms ${\bf A}_{1}({\bf X})\bdot\exd{\bf X} - \Phi_{1}({\bf X})\,c\,\exd t$, instead of the non-local terms $\langle{\bf A}_{1{\rm gc}}\rangle\bdot\exd{\bf X} - \langle\Phi_{1{\rm gc}}\rangle\,c\,\exd t$ used here. Despite this difference, the gyrocenter equations of motion derived in the present paper are gauge-independent in the sense that only perturbed electric and magnetic fields appear explicitly.

\subsection{Organization}

The remainder of the paper is organized as follows. In Sec.~\ref{sec:symplectic}, we derive a generic set of non-canonical gyrocenter Hamilton equations of motion based on the symplectic gyrocenter one-form \eqref{eq:Gamma_gy}. After introducing the symplectic gyrocenter transformation in Sec.~\ref{sec:symp_gyro}, we calculate the first-order gyroangle-averaged gyrocenter polarization displacement $\langle{\sf T}_{\rm gy}^{-1}({\bf X} + \vb{\rho}_{0})\rangle - {\bf X} = \epsilon\,\langle\vb{\rho}_{1{\rm gy}}\rangle$ and use it to find a suitable expression for the first-order gyrocenter symplectic momentum $\vb{\Pi}_{1{\rm gy}}$ so that gyrocenter polarization now appears solely in the gyrocenter symplectic one-form \eqref{eq:Gamma_gy}. The gyrocenter Euler-Lagrange and Hamilton equations are also presented. In addition, we derive the second-order gyrocenter Hamiltonian that is consistent with the first-order symplectic-momentum representation. We show that, in the zero-Larmor-radius limit, the second-order gyrocenter Hamilotnian depends only on the first-order perturbation electromagnetic fields $({\bf E}_{1},{\bf B}_{1})$. 

Next, using constrained variations for the gyrokinetic Vlasov-Maxwell dynamical fields,  the variational derivation of the symplectic gyrokinetic Vlasov-Maxwell equations is presented in Sec.~\ref{sec:symp_VP}, while the gyrokinetic energy-momentum and 
angular-momentum conservation laws are presented and proved explicitly in Sec.~\ref{sec:symp_laws}. Our work is summarized in Sec.~\ref{sec:Sum}, which includes a discussion of applications of a truncated set of symplectic gyrokinetic Vlasov-Maxwell equations in which the second-order gyrocenter Hamiltonian is omitted while still retaining the exact gyrokinetic conservation laws. Lastly, Bessel-function identities used in Sec.~\ref{sec:symp_gyro} are derived in App.~\ref{sec:Bessel}.

\section{\label{sec:symplectic}Gyrocenter Symplectic Structure}

The gyrocenter Hamilton equations of motion $\dot{\cal Z}^{a} = \{ {\cal Z}^{a},\; {\cal H}_{\rm gy}\}_{\rm gy}$ in eight-dimensional gyrocenter phase space are expressed in terms of the extended gyrocenter Hamiltonian ${\cal H}_{\rm gy} \equiv H_{\rm gy} - w$ and the extended gyrocenter Poisson bracket $\{\;,\;\}_{\rm gy}$, which are constructed by Lie-transform perturbation methods \citep{Brizard_Hahm_2007} directly from the gyrocenter phase-space transformation.

In order to construct the gyrocenter Poisson bracket $\{\;,\;\}_{\rm gy}$ from the gyrocenter extended one-form \eqref{eq:Gamma_gy}, we construct an $8\times 8$ Lagrange matrix $\vb{\omega}_{\rm gy}$ from the extended two-form $\omega_{\rm gy} = \exd\Gamma_{\rm gy}$ constructed as the exterior derivative of the gyrocenter extended one-form \eqref{eq:Gamma_gy}. From this matrix, we find the gyrocenter Jacobian
\begin{equation}
{\cal J}_{\rm gy} \;\equiv\; \sqrt{{\rm det}(\vb{\omega}_{\rm gy})} = \frac{e}{c}\,{\sf b}^{*}_{\rm gy}\bdot{\bf B}^{*}_{\rm gy} \;=\; \frac{e}{c}\,B_{\|{\rm gy}}^{*} \;=\; \frac{e}{c}\,B_{0} \;+\; {\sf b}^{*}_{\rm gy}\bdot\nabla\btimes\left(p_{\|}\,\bhat_{0} + \vb{\Pi}_{\rm gy}\right),
\label{eq:J_gy}
\end{equation}
where we use the definitions
\begin{equation}
\left. \begin{array}{rcl}
{\sf b}^{*}_{\rm gy} & \equiv & \bhat_{0} \;+\; \partial\vb{\Pi}_{\rm gy}/\partial p_{\|} \\
{\bf B}^{*}_{\rm gy} & \equiv & {\bf B}_{0} \;+\; (c/e)\,\nabla\btimes(p_{\|}\,\bhat_{0} + \vb{\Pi}_{\rm gy})
\end{array} \right\}.
\label{eq:bB_star}
\end{equation}
According to Eq.~\eqref{eq:scenarios}  [see Eq.~\eqref{eq:Pi_1gy}], the term $\partial\vb{\Pi}_{\rm gy}/\partial p_{\|}$ is perpendicular to ${\bf B}_{0}$, so that $\bhat_{0}\bdot{\sf b}^{*}_{\rm gy} = 1$.

We note that, because of the presence of $\vb{\Pi}_{\rm gy}$, the gyrocenter Jacobian \eqref{eq:J_gy} is time-dependent:
\begin{eqnarray}
\pd{{\cal J}_{\rm gy}}{t} &=& \pd{}{p_{\|}}\left(\pd{\vb{\Pi}_{\rm gy}}{t}\right)\bdot\frac{e}{c}\,{\bf B}^{*}_{\rm gy} \;+\; {\sf b}^{*}_{\rm gy}\bdot\nabla\btimes\pd{\vb{\Pi}_{\rm gy}}{t} \nonumber \\
 &=& \pd{}{p_{\|}}\left(\pd{\vb{\Pi}_{\rm gy}}{t}\bdot\frac{e}{c}\,{\bf B}^{*}_{\rm gy}\right) \;+\; \nabla\bdot\left(\pd{\vb{\Pi}_{\rm gy}}{t}\btimes{\sf b}^{*}_{\rm gy} \right),
  \label{eq:J_dot}
 \end{eqnarray}
where we used the relation
\begin{equation}
(e/c)\,\partial{\bf B}^{*}_{\rm gy}/\partial p_{\|} \;=\; \nabla\btimes{\sf b}^{*}_{\rm gy},
\label{eq:B_p||}
\end{equation}
which follows from the definitions \eqref{eq:bB_star}. 

\subsection{Symplectic Poisson bracket}

Assuming that the gyrocenter Jacobian ${\cal J}_{\rm gy}$ does not vanish (which is true under most general conditions), the Lagrange matrix $\vb{\omega}_{\rm gy}$ can be inverted, from which we construct the gyrocenter Poisson matrix ${\sf J}_{\rm gy}$, whose components $J_{\rm gy}^{ab} \equiv \{ {\cal Z}^{a}, {\cal Z}^{b}\}_{\rm gy}$ define the fundamental gyrocenter Poisson-bracket elements. Hence, according to this inversion procedure, we obtain the gyrocenter Poisson bracket
\begin{eqnarray}
\{ {\cal F},\; {\cal G}\}_{\rm gy} & = & \pd{\cal F}{w}\frac{\partial^{*}{\cal G}}{\partial t} - \frac{\partial^{*}{\cal F}}{\partial t}\pd{\cal G}{w} \;+\; \pd{\cal F}{\zeta}\pd{\cal G}{J} - \pd{\cal F}{J}\pd{\cal G}{\zeta} \nonumber \\
 & &+\; \frac{e{\bf B}^{*}_{\rm gy}}{c{\cal J}_{\rm gy}}\bdot\left(\nabla^{*}{\cal F}\pd{\cal G}{p_{\|}} - \pd{\cal F}{p_{\|}}\nabla^{*}{\cal G}\right) \;-\; \frac{{\sf b}^{*}_{\rm gy}}{{\cal J}_{\rm gy}}\bdot\nabla^{*}{\cal F}\btimes\nabla^{*}{\cal G},
\label{eq:PB_gy}
 \end{eqnarray}
 where the modified spatial gradient and time-derivative operators are defined as
\[ \nabla^{*} \;\equiv\; \nabla - \pd{\vb{\Pi}_{\rm gy}}{J}\;\pd{}{\zeta} - \pd{\vb{\Pi}_{\rm gy}}{t}\;\pd{}{w}  \;\;{\rm and}\;\;
\frac{\partial^{*}}{\partial t}  \;\equiv\; \pd{}{t} + \frac{{\sf b}^{*}_{\rm gy}}{{\cal J}_{\rm gy}}\bdot\left(\pd{\vb{\Pi}_{\rm gy}}{t}\btimes\pd{\vb{\Pi}_{\rm gy}}{J}\right)\;\pd{}{\zeta}. \]
Since the gyrocenter Poisson matrix satisfies the Liouville property
\begin{equation}
\pd{}{{\cal Z}^{a}} \left( {\cal J}_{\rm gy}\frac{}{} J_{\rm gy}^{ab}\right) \;=\; \pd{}{{\cal Z}^{a}} \left( {\cal J}_{\rm gy}\frac{}{} \{ {\cal Z}^{a}, {\cal Z}^{b}\}_{\rm gy} \right) \;=\; 0,
\label{eq:Liouville}
\end{equation}
the gyrocenter Poisson bracket \eqref{eq:PB_gy} may also be expressed in phase-space divergence form as
\begin{equation}
\{ {\cal F},\; {\cal G}\}_{\rm gy} \;\equiv\; \frac{1}{{\cal J}_{\rm gy}}\pd{}{{\cal Z}^{a}} \left( {\cal J}_{\rm gy}\,{\cal F}\frac{}{} \{ {\cal Z}^{a},\; {\cal G}\}_{\rm gy}\right).
\label{eq:PB_div}
\end{equation}
We note that the inversion procedure leading to the gyrocenter Poisson bracket \eqref{eq:PB_gy} guarantees the standard properties of Poisson brackets since the condition $\nabla\bdot{\bf B}^{*}_{\rm gy} = 0$ is satisfied.

\subsection{Symplectic gyrocenter Hamilton equations}

Using the Poisson bracket \eqref{eq:PB_gy}, we now write the gyrocenter Hamilton equations of motion $\dot{\cal Z}^{a} = \{ {\cal Z}^{a},\; {\cal H}_{\rm gy}\}_{\rm gy}$:
\begin{eqnarray}
\dot{\bf X}  & = & \frac{{\sf b}^{*}_{\rm gy}}{{\cal J}_{\rm gy}}\btimes\left(\nabla H_{\rm gy} + \pd{\vb{\Pi}_{\rm gy}}{t}\right) + \frac{e{\bf B}^{*}_{\rm gy}}{c{\cal J}_{\rm gy}}\;\pd{H_{\rm gy}}{p_{\|}}, \label{eq:X_dot} \\
\dot{p}_{\|} & = & -\;\frac{e{\bf B}^{*}_{\rm gy}}{c{\cal J}_{\rm gy}}\vb{\cdot}\left(\nabla H_{\rm gy} + \pd{\vb{\Pi}_{\rm gy}}{t}\right), \label{eq:p_dot}
\end{eqnarray}
where the extended gyrocenter Hamiltonian is defined as
\begin{equation}
{\cal H}_{\rm gy} = \frac{p_{\|}^{2}}{2m} + J\,\Omega + e\,\Psi_{\rm gy} - w \equiv H_{\rm gy} - w.
\label{eq:H_ext_gy}
\end{equation}
Here, the effective gyroangle-independent potential $\Psi_{\rm gy}$ depends on the perturbed electrostatic potential $\Phi_{1}$ and may depend on the perturbed vector potential ${\bf A}_{1}$ and magnetic field ${\bf B}_{1}$, depending on which representation is used. We note that, unless the Hamiltonian representation is chosen (for which $\vb{\Pi}_{1{\rm gy}} = 0$), the gyrocenter Hamilton equations \eqref{eq:X_dot}-\eqref{eq:p_dot} will contain explicit partial time derivatives $(\partial\vb{\Pi}_{1{\rm gy}}/\partial t \neq 0)$.

The remaining gyrocenter Hamilton equations are $\dot{J} = -\,\partial H_{\rm gy}/\partial\zeta \equiv 0$, which immediately follows from the gyroangle-independence of the gyrocenter Hamiltonian (see below), $\dot{\zeta} = \partial H_{\rm gy}/\partial J - \dot{\bf X}\bdot\partial\vb{\Pi}_{\rm gy}/\partial J$, $\dot{t} = -\,\partial{\cal H}_{\rm gy}/\partial w = 1$, and $\dot{w} = \partial H_{\rm gy}/\partial t - \dot{\bf X}\bdot\partial\vb{\Pi}_{\rm gy}/\partial t$. Lastly, because of the Liouville property \eqref{eq:Liouville}, the gyrocenter Hamilton equations \eqref{eq:X_dot}-\eqref{eq:p_dot} satisfy the gyrocenter Liouville Theorem:
\begin{equation}
\pd{{\cal J}_{\rm gy}}{t} + \nabla\bdot\left({\cal J}_{\rm gy}\frac{}{}\dot{\bf X}\right) + \pd{}{p_{\|}}\left({\cal J}_{\rm gy}\frac{}{}\dot{p}_{\|}\right) \;=\; 0,
\label{eq:Liouville_id}
\end{equation}
where
\begin{eqnarray*}
\nabla\bdot\left({\cal J}_{\rm gy}\frac{}{}\dot{\bf X}\right) & = & \nabla\btimes{\sf b}^{*}_{\rm gy}\bdot\nabla H_{\rm gy} \;+\; \nabla\bdot\left({\sf b}^{*}_{\rm gy}\btimes\pd{\vb{\Pi}_{\rm gy}}{t}\right) \;+\; \frac{e}{c}\;{\bf B}^{*}_{\rm gy}\bdot\nabla\left(\pd{H_{\rm gy}}{p_{\|}}\right), \\
 \pd{}{p_{\|}}\left({\cal J}_{\rm gy}\frac{}{}\dot{p}_{\|}\right) & = & -\;\frac{e}{c}\pd{{\bf B}^{*}_{\rm gy}}{p_{\|}}\bdot\nabla H_{\rm gy} \;-\; \frac{e}{c}\;{\bf B}^{*}_{\rm gy}\bdot\nabla\left(\pd{H_{\rm gy}}{p_{\|}}\right) \;-\; \pd{}{p_{\|}}\left(\frac{e}{c}{\bf B}^{*}_{\rm gy}\bdot\pd{\vb{\Pi}_{\rm gy}}{t}\right),
\end{eqnarray*}
from which we recover Eq.~\eqref{eq:J_dot} when Eq.~\eqref{eq:B_p||} is used.

\section{\label{sec:symp_gyro}Symplectic gyrocenter phase-space transformation}

Using the phase-space Lagrangian Lie-transform perturbation methods \citep{Littlejohn_1982,Littlejohn_1983}, the derivations of the gyrocenter symplectic one-form \eqref{eq:Gamma_gy} and the gyrocenter Hamiltonian \eqref{eq:H_ext_gy} proceed by a near-identity phase-space transformation (associated with a small parameter $\epsilon$ that denotes the amplitude of the first-order perturbation fields) from the perturbed symplectic guiding-center one-form
\begin{eqnarray}
\Gamma_{\rm gc} &=& \frac{e}{c}\,\left[ {\bf A}_{0}^{*}\bdot\exd{\bf X}_{0} \;+\frac{}{} \epsilon\,{\bf A}_{1{\rm gc}}\bdot\exd({\bf X}_{0} + \vb{\rho}_{0}) \right] + J_{0}\,\exd\zeta_{0} - w_{0}\,\exd t \nonumber \\
 &\equiv& \Gamma_{0{\rm gc}} \;+\; \epsilon\,\frac{e}{c}\,{\bf A}_{1{\rm gc}}\bdot\exd({\bf X}_{0} + \vb{\rho}_{0}),
\label{eq:Gamma_gc}
\end{eqnarray}
and the perturbed guiding-center Hamiltonian
\begin{equation}
{\cal H}_{\rm gc} = \frac{p_{\|0}^{2}}{2m} + J_{0}\,\Omega_{0} + \epsilon\;e\,\Phi_{1{\rm gc}} - w_{0} \;\equiv\; {\cal H}_{0{\rm gc}} + \epsilon e\,\Phi_{1{\rm gc}},
\label{eq:H_gc}
\end{equation}
where  the guiding-center coordinates are ${\cal Z}_{0}^{a} = ({\bf X}_{0},p_{\|0},J_{0},\zeta_{0},w_{0},t)$ and the term 
\[ (e/c)\,{\bf A}_{0}^{*} \equiv (e/c)\,{\bf A}_{0} + p_{0\|}\,\bhat_{0} - J_{0} ({\bf R} + \frac{1}{2}\,\nabla\btimes\bhat_{0}) \] 
includes gyrogauge and higher-order guiding-center corrections \citep{Tronko_Brizard_2015}. We now look for the gyrocenter phase-space coordinates constructed as expansions in powers of $\epsilon$:
 \begin{equation}
 {\cal Z}^{a} = {\cal Z}_{0}^{a} + \epsilon\,{\cal G}_{1}^{a} + \epsilon^{2} \left( {\cal G}_{2}^{a} + \frac{1}{2}{\cal G}_{1}^{b}\pd{{\cal G}_{1}^{a}}{{\cal Z}_{0}^{b}}\right) + \cdots,
 \label{eq:ovZ_Z}
 \end{equation}
where the generating vector field ${\cal G}_{n}^{a}$ generates the gyrocenter transformation at $n$th-order. The fact that the dimensionless parameter $\epsilon \ll 1$ is considered small implies that the transformation \eqref{eq:ovZ_Z} is a near-identity transformation that is invertible:
 \begin{equation}
 {\cal Z}_{0}^{a} = {\cal Z}^{a} - \epsilon\,{\cal G}_{1}^{a} - \epsilon^{2} \left( {\cal G}_{2}^{a} - \frac{1}{2}{\cal G}_{1}^{b}\pd{{\cal G}_{1}^{a}}{{\cal Z}^{b}}\right) + \cdots.
 \label{eq:Z_ovZ}
 \end{equation}
 We note that the Jacobian ${\cal J}_{\rm gy}$ can be constructed from the unperturbed (guiding-center) Jacobian ${\cal J}_{0{\rm gy}} = {\cal J}_{\rm gc}$ as ${\cal J}_{\rm gy} \equiv {\cal J}_{0{\rm gy}} - \epsilon\,\partial_{a}({\cal J}_{0{\rm gy}}\,{\cal G}_{1}^{a}) + \cdots$ \citep{AJB_2016}.

The gyrocenter symplectic one-form \eqref{eq:Gamma_gy} is constructed from the perturbed symplectic guiding-center one-form \eqref{eq:Gamma_gc} by Lie-transform method \citep{Brizard_Hahm_2007}:
\begin{equation} 
\Gamma_{\rm gy} \;\equiv\; {\sf T}_{\rm gy}^{-1}\Gamma_{\rm gc} \;+\; \exd S,
 \label{eq:Gamma_Lie}
 \end{equation}
where the gyrocenter push-forward operator ${\sf T}_{\rm gy}^{-1} \equiv \cdots \exp(-\epsilon^{2}\pounds_{2})\,\exp(-\epsilon\pounds_{1})$ is defined in terms of Lie derivatives $(\pounds_{1},\pounds_{2},\cdots)$ that are generated by the generating vector fields $({\cal G}_{1},{\cal G}_{2},\cdots)$ and the gauge function $S \equiv \epsilon\,S_{1} + \epsilon^{2}S_{2} + \cdots$ represents the generating function for the canonical part of the gyrocenter phase-space transformation. 
 
 Once the generating vector fields $({\cal G}_{1},{\cal G}_{2},\cdots)$ are obtained from the solution of Eq.~\eqref{eq:Gamma_Lie}, the gauge functions $(S_{1},S_{2},\cdots)$ are determined from the solution of the gyrocenter Hamiltonian
\begin{equation}
{\cal H}_{\rm gy} \;\equiv\; {\sf T}_{\rm gy}^{-1}{\cal H}_{\rm gc} \;=\; \epsilon\,e\,\langle\Phi_{1{\rm gc}}\rangle \;+\; K_{\rm gy} \;-\; w,
\label{eq:H_Lie}
\end{equation}
where the gyrocenter kinetic energy $K_{\rm gy} = p_{\|}^{2}/2m + \mu\,B_{0} + \epsilon\,K_{1{\rm gy}} + \epsilon^{2}K_{2{\rm gy}} + \cdots$ is gyroangle-independent up to an arbitrary order in $\epsilon$, which therefore guarantees the exact invariance of the gyrocenter gyroaction $J$ (even though it is an adiabatic invariant of the exact particle dynamics).

\subsection{First-order gyrocenter analysis}

At first order in the perturbation analysis \eqref{eq:Gamma_Lie}, we find the first-order symplectic equation
\[ \vb{\Pi}_{1{\rm gy}}\bdot\pd{\bf X}{{\cal Z}^{b}} \;=\; \frac{e}{c}\,{\bf A}_{1{\rm gc}}\bdot\pd{({\bf X} + \vb{\rho}_{0})}{{\cal Z}^{b}} \;-\; {\cal G}_{1}^{a}\omega_{0ab} \;+\; \pd{S_{1}}{{\cal Z}^{b}}, \]
from which we obtain the first-order components
\begin{equation}
{\cal G}_{1}^{a} \;\equiv\; \{ S_{1}, {\cal Z}^{a}\}_{0} \;+\; \frac{e}{c}\,{\bf A}_{1{\rm gc}}\bdot\left\{ {\bf X} + \vb{\rho}_{0},\; {\cal Z}^{a}\right\}_{0} \;-\; \vb{\Pi}_{1{\rm gy}}\bdot\{ {\bf X}, {\cal Z}^{a}\}_{0},
\label{eq:G1_a}
\end{equation}
where $\{,\}_{0}$ denotes the unperturbed gyrocenter (guiding-center) Poisson bracket \eqref{eq:PB_gy} (with $\epsilon = 0$), obtained by inverting the unperturbed Lagrange matrix $\omega_{0ab}$. The contributions in the first-order gyrocenter phase-space transformation generated by Eq.~\eqref{eq:G1_a} include a canonical part (generated by $S_{1}$) and the non-canonical substitution of the gyroangle-independent symplectic momentum $\vb{\Pi}_{1{\rm gy}}$ after ${\bf A}_{1{\rm gc}}\bdot\exd({\bf X}_{0} + \vb{\rho}_{0})$ is removed  from the perturbed symplectic structure \eqref{eq:Gamma_gc}.

The canonical gauge function $S_{1}$ is determined from the first-order Hamilton equation
\begin{equation}
e\,\Psi_{1{\rm gy}} \;=\; e\,\psi_{1{\rm gc}} \;+\; \vb{\Pi}_{1{\rm gy}}\bdot\dot{\bf X}_{0} \;-\; \{ S_{1}, {\cal H}_{0} \}_{0}, 
\label{eq:S1_def}
\end{equation}
where $\dot{\bf X}_{0} \equiv \{ {\bf X}, {\cal H}_{0}\}_{0} = (p_{\|}/m)\,\bhat_{0}$ is the lowest-order unperturbed gyrocenter velocity, and the effective first-order perturbation potential $\psi_{1{\rm gc}} \equiv 
\Phi_{1{\rm gc}} - {\bf A}_{1{\rm gc}}\bdot{\bf v}_{0}/c$ is expressed in terms of the lowest-order particle velocity ${\bf v}_{0} = (p_{\|}/m)\,\bhat_{0} + \Omega\,\partial\vb{\rho}_{0}/\partial\zeta$. Since we want 
$\Psi_{1{\rm gy}}$ to be gyroangle-independent, it is defined as the gyroangle-averaged part of the right side of Eq.~\eqref{eq:S1_def}:
\begin{equation}
e\,\Psi_{1{\rm gy}} \;=\; e\,\langle\psi_{1{\rm gc}}\rangle \;+\; \frac{p_{\|}}{m}\;\vb{\Pi}_{1{\rm gy}}\bdot\bhat_{0},
\label{eq:Psi_1}
\end{equation}
where $S_{1}$ is assumed to be explicitly gyroangle-dependent, which is determined from the remaining gyroangle-dependent terms in Eq.~\eqref{eq:S1_def}:
\begin{equation}
\{ S_{1}, {\cal H}_{0} \}_{0} \;=\; e\,\wt{\psi}_{1{\rm gc}} \;\equiv\; e\,\left(\psi_{1{\rm gc}} \;-\frac{}{} \langle\psi_{1{\rm gc}}\rangle\right),
\label{eq:S1_dot}
\end{equation}
where, by definition, we took $\vb{\Pi}_{1{\rm gy}}$ to be gyroangle-independent. 

We note that, to lowest order in the standard gyrokinetic ordering, $\{ S_{1}, {\cal H}_{0} \}_{0} \simeq \Omega\,\partial S_{1}/\partial\zeta$, so that $S_{1}$ can be explicitly obtained in terms of indefinite gyroangle integrals of the right side of Eq.~\eqref{eq:S1_dot}. In addition, we note that the definition of the gyrocenter gauge function $S_{1}$ is independent of the choice of the gyrocenter symplectic momentum $\vb{\Pi}_{\rm gy}$. Hence, the definition of the gyrocenter gyroaction $J$ is also independent of that choice. This equivalence of representations was discussed previously in the context of guiding-center theory \citep{Tronko_Brizard_2015} and gyrocenter theory \citep{Brizard_2017b}.

\subsection{Symplectic gyrocenter polarization displacement}

We now choose the gyrocenter symplectic momentum $\vb{\Pi}_{\rm gy}$ such that the gyrocenter polarization displacement yields the standard first-order gyrocenter polarization. In order to calculate this gyrocenter polarization displacement, we begin with the first-order gyrocenter transformation \eqref{eq:G1_a}, from which we calculate the gyrocenter displacement \citep{Brizard_2013}
\begin{equation}
\vb{\rho}_{1{\rm gy}} \;\equiv\; -\,{\cal G}_{1}\cdot\exd({\bf X} + \vb{\rho}_{0}) \;=\; \{ {\bf X} + \vb{\rho}_{0},\; S_{1}\}_{0} \;+\; \vb{\Pi}_{1{\rm gy}}\btimes\frac{c\bhat_{0}}{eB_{0}},
\label{eq:rho_1}
\end{equation}
where the contribution from ${\bf A}_{1{\rm gc}}$ cancels out because of the identity $\{ {\bf X} + \vb{\rho}_{0}, {\bf X} + \vb{\rho}_{0}\}_{0} \equiv 0$. Next, we calculate the gyroangle-averaged first-order gyrocenter displacement
\begin{equation}
\langle\vb{\rho}_{1{\rm gy}}\rangle \;=\; \left\langle\{\vb{\rho}_{0},\frac{}{} S_{1}\}_{0}\right\rangle \;+\; \vb{\Pi}_{1{\rm gy}}\btimes\frac{c\bhat_{0}}{eB_{0}} \;\simeq\; -\,\pd{}{J}\left\langle\vb{\rho}_{0}\,\pd{S_{1}}{\zeta}\right\rangle
\;+\; \vb{\Pi}_{1{\rm gy}}\btimes\frac{c\bhat_{0}}{eB_{0}},
\label{eq:rho1_av}
\end{equation}
where the Poisson-bracket term is evaluated to lowest order in the guiding-center and gyrokinetic orderings \citep{Brizard_Hahm_2007}, i.e., we assume that the background magnetic field is uniform in the calculations that follow.

If we now use the lowest-order solution to Eq.~\eqref{eq:S1_dot}, which assumes that the background magnetic field is uniform, we find
\begin{eqnarray} 
\left\langle\vb{\rho}_{0}\,\pd{S_{1}}{\zeta}\right\rangle & = & \frac{e}{\Omega_{0}} \left\langle\vb{\rho}_{0}\left( \Phi_{1{\rm gc}} - \frac{v_{\|}}{c}\,A_{1\|{\rm gc}}\right)\right\rangle \;-\; \frac{e}{c}\,\left\langle\vb{\rho}_{0}\;\vb{\rho}_{0}\bdot\bhat_{0}\btimes{\bf A}_{1{\rm gc}}\right\rangle \nonumber \\
 & \equiv & \frac{e}{\Omega_{0}}\,\left\langle\vb{\rho}_{0}\;{\sf T}_{\rm gc}^{-1}\right\rangle \left( \Phi_{1} - \frac{v_{\|}}{c}\,A_{1\|}\right) \;-\; \frac{e}{c}\,\left\langle\vb{\rho}_{0}\;\vb{\rho}_{0}\frac{}{} {\sf T}_{\rm gc}^{-1}\right\rangle\bdot\bhat_{0}\btimes{\bf A}_{1}.
  \label{eq:A1}
\end{eqnarray}
In order to find an explicit expression for Eqs.~\eqref{eq:rho1_av}-\eqref{eq:A1}, we need to evaluate the gyroangle-averaged operators $\langle\vb{\rho}_{0}{\sf T}_{\rm gc}^{-1}\rangle$ and $\langle\vb{\rho}_{0}\vb{\rho}_{0}{\sf T}_{\rm gc}^{-1}\rangle$, as well as their derivatives, which are computed in App.~\ref{sec:Bessel}. By replacing the first term in Eq.~\eqref{eq:rho1_av} with Eq.~\eqref{eq:App_id}:
\begin{eqnarray}
\pd{}{J}\left\langle\vb{\rho}_{0}\,\pd{S_{1}}{\zeta}\right\rangle  & = & -\,\frac{e}{m\Omega_{0}^{2}}\left(\langle{\bf E}_{1\bot{\rm gc}}\rangle + \frac{p_{\|}\bhat_{0}}{mc}\btimes\langle{\bf B}_{1\bot{\rm gc}}\rangle\right) \nonumber \\
 &&+\; \frac{\mu}{m\Omega_{0}^{2}}\;\nabla_{\bot}\langle\langle B_{1\|{\rm gc}}\rangle\rangle - \frac{\bhat_{0}}{B_{0}}\btimes\langle{\bf A}_{1\bot{\rm gc}}\rangle,
\label{eq:Bessel_gy}
\end{eqnarray}
 where the symbol $\langle\langle \cdots\rangle\rangle$ is introduced \cite{Porazik_Lin_2011} to denote a gyro-surface average (see App.~\ref{sec:Bessel}), we obtain the first-order gyroangle-averaged gyrocenter displacement
\begin{eqnarray}
\langle\vb{\rho}_{1{\rm gy}}\rangle &=& \frac{e}{m\Omega_{0}^{2}}\left( \langle{\bf E}_{1\bot{\rm gc}}\rangle + \frac{p_{\|}\bhat_{0}}{mc}\btimes\langle{\bf B}_{1\bot{\rm gc}}\rangle\right) \nonumber \\
 &&+ \left( \vb{\Pi}_{1{\rm gy}} - \frac{e}{c}\,\langle{\bf A}_{1{\rm gc}}\rangle\right) \times\frac{\bhat_{0}}{m\Omega_{0}} -  \frac{\mu}{m\Omega_{0}^{2}}\;\nabla_{\bot}\langle\langle B_{1\|{\rm gc}}\rangle\rangle.
\label{eq:rho1_gy} 
\end{eqnarray}
In the present work, we choose the first-order gyroangle-averaged gyrocenter displacement
\begin{equation}
\langle\vb{\rho}_{1{\rm gy}}\rangle \;\equiv\; -\;\frac{\mu}{m\Omega_{0}^{2}}\;\nabla_{\bot}\langle\langle B_{1\|{\rm gc}}\rangle\rangle,
\label{eq:rho_1gy}
\end{equation}
which is the first-order correction to the zeroth-order (guiding-center) polarization displacement \citep{Brizard_2013, Tronko_Brizard_2015}
\begin{equation}
\langle\vb{\rho}_{0{\rm gy}}\rangle - \nabla\bdot\left\langle\frac{1}{2}\,\vb{\rho}_{0{\rm gy}}\vb{\rho}_{0{\rm gy}}\right\rangle \;=\; \frac{\bhat_{0}}{\Omega_{0}}\btimes\dot{\bf X}_{\rm gc} \;=\; -\;\frac{1}{m\Omega_{0}^{2}} \left( \mu\;\nabla_{\bot}B_{0} \;+\; 
\frac{p_{\|}^{2}}{m}\,\bhat_{0}\bdot\nabla\bhat_{0} \right).
\end{equation}
Hence, from Eq.\eqref{eq:rho_1}, we choose the first-order gyrocenter symplectic momentum as
\begin{equation}
\vb{\Pi}_{1{\rm gy}} \;=\; \frac{e}{c}\,\langle{\bf A}_{1{\rm gc}}\rangle \;+\; \left(\langle{\bf E}_{1{\rm gc}}\rangle + \frac{p_{\|}\bhat_{0}}{mc}\btimes\langle{\bf B}_{1{\rm gc}}\rangle\right)\btimes\frac{e\bhat_{0}}{\Omega_{0}} \;\equiv\; 
\frac{e}{c}\,\langle{\bf A}_{1{\rm gc}}\rangle \;+\; {\bf P}_{1{\rm gy}},
\label{eq:Pi_1gy}
\end{equation}
which incorporates all three scenarios introduced in Eq.~\eqref{eq:scenarios}. In addition, the first-order gyrocenter Hamiltonian \eqref{eq:Psi_1} becomes
\begin{equation}
e\,\Psi_{1{\rm gy}} \;=\; e\,\langle\Phi_{1{\rm gc}}\rangle \;+\; \mu\;\langle\langle B_{1\|{\rm gc}}\rangle\rangle.
\label{eq:Psi1_gy}
\end{equation}
Note that Eqs.~\eqref{eq:Pi_1gy}-\eqref{eq:Psi1_gy} depend explicitly on the perturbed electric and magnetic fields $({\bf E}_{1},{\bf B}_{1})$ as well as the minimal-coupling combination of the perturbed electromagnetic potentials $(\Phi_{1},{\bf A}_{1})$, which guarantees gauge invariance in the zero-Larmor-radius limit.

\subsection{Second-order Gyrocenter Analysis}

Next, in order to calculate the second-order gyrocenter Hamiltonian in the present symplectic representation, we perform the second-order gyrocenter analysis leading to second-order corrections to the gyrocenter Hamiltonian and Poisson bracket. We begin with the second-order gyrocenter symplectic one-form
\begin{equation}
\Gamma_{{\rm gy}2} \;=\; -\,\pounds_{2}\Gamma_{{\rm gc}0} \;-\; \frac{1}{2}\,\pounds_{1}\left(\Gamma_{{\rm gc}1} + \Gamma_{{\rm gy}1}\right) \;+\; \exd S_{2} \;\equiv\; 0,
\label{eq:Gamma_2}
\end{equation}
which is chosen to be unperturbed at the second order ($\Gamma_{{\rm gy}2} = 0$), with
\begin{eqnarray*}
\pounds_{2}\Gamma_{{\rm gc}0} & = & {\cal G}_{2}^{a}\,\omega_{0ab}\,\exd {\cal Z}^{b}, \\
\pounds_{1}\Gamma_{{\rm gc}1} & = & \vb{\rho}_{1{\rm gy}}\bdot\left[\frac{e}{c} {\bf B}_{1{\rm gc}}\btimes\exd({\bf X} + \vb{\rho}_{0}) \;+\; \frac{e}{c}\;\pd{{\bf A}_{1{\rm gc}}}{t}\;\exd t\right], \\
\pounds_{1}\Gamma_{{\rm gy}1} & = & -\,{\cal G}_{1}^{\bf X}\bdot\left[ (\nabla\btimes\vb{\Pi}_{1{\rm gy}})\btimes\exd{\bf X} + \pd{\vb{\Pi}_{1{\rm gy}}}{p_{\|}}\;\exd p_{\|} + \pd{\vb{\Pi}_{1{\rm gy}}}{\mu}\;\exd\mu + \pd{\vb{\Pi}_{1{\rm gy}}}{t}\;\exd t \right] \\
 &&+\; \left( {\cal G}_{1}^{p_{\|}}\;\pd{\vb{\Pi}_{1{\rm gy}}}{p_{\|}} + {\cal G}_{1}^{\mu}\;\pd{\vb{\Pi}_{1{\rm gy}}}{\mu}\right)\bdot\exd{\bf X},
\end{eqnarray*}
where ${\cal G}_{1}^{a}$ and $\vb{\rho}_{1{\rm gy}}$ are given by Eqs.~\eqref{eq:G1_a} and \eqref{eq:rho_1}, respectively. The inversion of the zeroth-order Lagrange bracket $\omega_{0ab}$ in the first term on the right side of Eq.~\eqref{eq:Gamma_2} yields the solution for the second-order generating vector field
\begin{eqnarray}
{\cal G}_{2}^{a} & = & \{ S_{2}, {\cal Z}^{a}\}_{0} \;+\; \frac{e}{2c}\,\vb{\rho}_{1{\rm gy}}\bdot\left(\{ {\bf X} + \vb{\rho}_{0}, {\cal Z}^{a}\}_{0}\btimes{\bf B}_{1{\rm gc}} \;-\; \pd{{\bf A}_{1{\rm gc}}}{t}\;\{t, {\cal Z}^{a}\}_{0}\right)
\nonumber \\
 &  &-\; \frac{1}{2} \left((\nabla\btimes\vb{\Pi}_{1{\rm gy}})\btimes{\cal G}_{1}^{\bf X} + {\cal G}_{1}^{p_{\|}}\;\pd{\vb{\Pi}_{1{\rm gy}}}{p_{\|}} + {\cal G}_{1}^{\mu}\;\pd{\vb{\Pi}_{1{\rm gy}}}{\mu}\right) \bdot\{ {\bf X}, {\cal Z}^{a}\}_{0} \nonumber \\
  &  &+\; \frac{1}{2} \left(  \pd{\vb{\Pi}_{1{\rm gy}}}{p_{\|}}\;\{ p_{\|}, {\cal Z}^{a}\}_{0} + \pd{\vb{\Pi}_{1{\rm gy}}}{\mu}\;\{ \mu, {\cal Z}^{a}\}_{0} + \pd{\vb{\Pi}_{1{\rm gy}}}{t}\;\{ t, {\cal Z}^{a}\}_{0}\right)\bdot{\cal G}_{1}^{\bf X}.
  \label{eq:G2_a}
\end{eqnarray}
Using the first-order and second-order generating vector fields \eqref{eq:G1_a} and \eqref{eq:G2_a}, we can now derive the second-order Hamiltonian equation
\begin{eqnarray}
{\cal H}_{2{\rm gy}} & = & -\;{\cal G}_{2}^{a}\;\pd{{\cal H}_{0{\rm gy}}}{{\cal Z}^{a}} \;-\; \frac{1}{2}\;{\cal G}_{1}^{a}\left( \pd{{\cal H}_{1{\rm gc}}}{{\cal Z}^{a}} + \pd{{\cal H}_{1{\rm gy}}}{{\cal Z}^{a}} \right) \nonumber \\
    & = & -\;\{ S_{2}, {\cal H}_{0{\rm gy}} \}_{0} \;-\; \frac{e}{2}\,\vb{\rho}_{1{\rm gy}}\bdot\left({\bf E}_{1{\rm gc}} \;+\; \frac{{\bf v}_{0}}{c}\btimes{\bf B}_{1{\rm gc}} \right) \nonumber \\
     &&+\; \frac{e}{2} \,{\cal G}_{1}^{\bf X}\bdot\left\langle {\bf E}_{1{\rm gc}} \;+\; \frac{{\bf v}_{0}}{c}\btimes{\bf B}_{1{\rm gc}} \right\rangle \;-\; \frac{1}{2}\,{\cal G}_{1}^{\bf X}\bdot\frac{d_{0}}{dt}\left(\vb{\Pi}_{1{\rm gy}} \;-\; \frac{e}{c}\,\langle{\bf A}_{1{\rm gc}}\rangle\right) \nonumber \\
     &&-\; \frac{1}{2}\,{\cal G}_{1}^{p_{\|}}\,\left(\vb{\Pi}_{1{\rm gy}} \;-\; \frac{e}{c}\,\langle{\bf A}_{1{\rm gc}}\rangle\right)\bdot\frac{\bhat_{0}}{m} \;-\; \frac{e}{2}\,{\cal G}_{1}^{\mu}\;
      \pd{\langle\psi_{1{\rm gc}}\rangle}{\mu},
   \label{eq:H2_gy_EB}
 \end{eqnarray}
where ${\bf v}_{0} = p_{\|}\bhat_{0}/m + \Omega_{0}\,\partial\vb{\rho}_{0}/\partial\zeta$ and $\psi_{1{\rm gc}} = \Phi_{1{\rm gc}} - {\bf A}_{1{\rm gc}}\bdot{\bf v}_{0}/c \equiv \langle\psi_{1{\rm gc}}\rangle + \wt{\psi}_{1{\rm gc}}$, with $e\,\Psi_{1{\rm gy}} = e\,\langle\psi_{1{\rm gc}}\rangle + \vb{\Pi}_{1{\rm gy}}\bdot\dot{\bf X}_{0}$. In addition, we used the identity
\begin{equation}
{\bf E}_{1{\rm gc}} \;+\; \frac{{\bf v}_{0}}{c} \btimes{\bf B}_{1{\rm gc}} \;=\; -\,\nabla\psi_{1{\rm gc}} \;-\; \frac{1}{c}\,\frac{d{\bf A}_{1{\rm gc}}}{dt},
\label{eq:EB_psi}
\end{equation}
and the definitions $d/dt = \partial/\partial t + {\bf v}_{0}\bdot\nabla$, $d_{0}/dt \equiv \{\;,\;{\cal H}_{0}\}_{0} = \partial/\partial t + \dot{\bf X}_{0}\bdot\nabla$ and $\{{\bf X} + \vb{\rho}_{0},\; \Phi_{1{\rm gc}}\}_{0} \equiv 0$. By convention, the second-order gyrocenter Hamiltonian is defined as the gyroangle-averaged part of the right side of Eq.~\eqref{eq:H2_gy_EB}:
\begin{eqnarray}
{\cal H}_{2{\rm gy}} & = & -\; \frac{e}{2}\,\left\langle\vb{\rho}_{1{\rm gy}}\bdot\left({\bf E}_{1{\rm gc}} \;+\; \frac{{\bf v}_{0}}{c}\btimes{\bf B}_{1{\rm gc}} \right)\right\rangle \;+\; \frac{e}{2} \,\left\langle{\cal G}_{1}^{\bf X}\right\rangle\bdot\left\langle {\bf E}_{1{\rm gc}} \;+\; \frac{{\bf v}_{0}}{c}\btimes{\bf B}_{1{\rm gc}} \right\rangle \nonumber \\
    &&-\; \frac{1}{2}\,\left\langle{\cal G}_{1}^{\bf X}\right\rangle\bdot\frac{d_{0}}{dt}\left(\vb{\Pi}_{1{\rm gy}} \;-\; \frac{e}{c}\,\langle{\bf A}_{1{\rm gc}}\rangle\right) \;-\; \frac{1}{2}\,\left\langle{\cal G}_{1}^{p_{\|}}\right\rangle\,\left(\vb{\Pi}_{1{\rm gy}} \;-\; \frac{e}{c}\,\langle{\bf A}_{1{\rm gc}}\rangle\right)\bdot\frac{\bhat_{0}}{m} \nonumber \\
     &&-\; \frac{e}{2}\,\left\langle{\cal G}_{1}^{\mu}\right\rangle\;\pd{\langle\psi_{1{\rm gc}}\rangle}{\mu},
 \label{eq:H2_gy_final}
 \end{eqnarray} 
where 
\begin{eqnarray*}
\left\langle{\cal G}_{1}^{\bf X}\right\rangle & = & \frac{\bhat_{0}}{m\Omega_{0}}\btimes\left(\vb{\Pi}_{1{\rm gy}} \;-\; \frac{e}{c}\,\langle{\bf A}_{1{\rm gc}}\rangle\right) \;=\; \frac{\bhat_{0}}{m\Omega_{0}}\btimes\left[\left(\langle{\bf E}_{1{\rm gc}}\rangle \;+\;
\frac{p_{\|}\bhat_{0}}{mc}\btimes\langle{\bf B}_{1{\rm gc}}\rangle\right)\btimes\frac{e\bhat_{0}}{\Omega_{0}}\right], \\
\left\langle{\cal G}_{1}^{p_{\|}}\right\rangle & = & -\;\bhat_{0}\bdot\left(\vb{\Pi}_{1{\rm gy}} \;-\; \frac{e}{c}\,\langle{\bf A}_{1{\rm gc}}\rangle\right) \;=\; 0, \\
\left\langle{\cal G}_{1}^{\mu}\right\rangle & = & \frac{e\Omega_{0}}{cB_{0}}\left\langle{\bf A}_{1{\rm gc}}\bdot\pd{\vb{\rho}_{0}}{\zeta}\right\rangle \;=\; -\;\frac{\mu}{B_{0}}\;\langle\langle B_{1\|{\rm gc}}\rangle\rangle.
\end{eqnarray*}
In what follows, we verify that we can recover known results from Eq.~\eqref{eq:H2_gy_final} before deriving the second-order gyrocenter Hamiltonian in the zero-Larmor-radius (ZLR) limit.

\subsubsection{Second-order Hamiltonian representation}

We would like to verify that Eq.~\eqref{eq:H2_gy_final} yields the standard result in the Hamiltonian representation, in which $\vb{\Pi}_{1{\rm gy}} = 0$ and $\vb{\rho}_{1{\rm gy}} = -\,\{ S_{1}, {\bf X} + \vb{\rho}_{0}\}_{0}$, so that Eq.~\eqref{eq:H2_gy_final} becomes
\begin{equation}
{\cal H}_{2{\rm gy}}^{\rm Ham} \;=\; -\; \frac{e}{2}\, \left\langle\vb{\rho}_{1{\rm gy}}\bdot\left({\bf E}_{1{\rm gc}} \;+\; \frac{{\bf v}_{0}}{c}\btimes{\bf B}_{1{\rm gc}} \right)\right\rangle \;-\;  \frac{e}{2} \left\langle{\cal G}_{1}^{a}\right\rangle
\pd{\langle\psi_{1{\rm gc}}\rangle}{{\cal Z}^{a}},
\label{eq:H_ham}
\end{equation}
where we have omitted explicit time derivatives. First, using the identity \eqref{eq:EB_psi}, the first term becomes
\begin{eqnarray*}
\left\langle \vb{\rho}_{1{\rm gy}}\bdot\left( {\bf E}_{1{\rm gc}} + \frac{{\bf v}_{0}}{c}\btimes{\bf B}_{1{\rm gc}}\right)\right\rangle & = & \left\langle \{ S_{1},\;\psi_{1{\rm gc}}\}_{0} \;+\; \{ S_{1},\; {\bf v}_{0}\}_{0}\bdot\frac{1}{c}{\bf A}_{1{\rm gc}}
\right\rangle \nonumber \\
 &&+\; \frac{1}{c}\frac{d_{0}}{dt}\left\langle \{ S_{1},\frac{}{} {\bf X} + \vb{\rho}_{0}\}_{0}\bdot{\bf A}_{1{\rm gc}}\right\rangle \\
 &&-\; \frac{1}{c} \left\langle \{S_{1},\;{\bf v}_{0}\}_{0}\bdot{\bf A}_{1{\rm gc}} \;+\; \left\{ \frac{d_{0}S_{1}}{dt},\; {\bf X} + \vb{\rho}_{0}\right\}_{0}
 \bdot{\bf A}_{1{\rm gc}} \right\rangle \\
  &=& \left\langle \left\{ S_{1},\frac{}{} \psi_{1{\rm gc}}\right\}_{0}\right\rangle \;+\; \frac{1}{c} \left\langle {\bf A}_{1{\rm gc}}\bdot\left\{ {\bf X} + \vb{\rho}_{0},\frac{}{} e\,\wt{\psi}_{1{\rm gc}}\right\}_{0}\right\rangle,
\end{eqnarray*}
where exact time derivatives can be removed from the expression of any Hamiltonian, while the last terms in Eq.~\eqref{eq:H_ham} yield
\begin{eqnarray*} 
-\,\frac{e}{2} \left\langle{\cal G}_{1}^{a}\right\rangle\pd{\langle\psi_{1{\rm gc}}\rangle}{{\cal Z}^{a}} & = & 
-\,\frac{e}{2c} \left\langle {\bf A}_{1{\rm gc}}\bdot\left\{{\bf X} + \vb{\rho}_{0},\frac{}{} e\,\langle\psi_{1{\rm gc}}\rangle\right\}_{0} \right\rangle \\
 &=&  -\,\frac{e}{2c} \left\langle {\bf A}_{1{\rm gc}}\bdot\left\{{\bf X} + \vb{\rho}_{0},\frac{}{} e\,\left(\psi_{1{\rm gc}} - \wt{\psi}_{1{\rm gc}}\right)\right\}_{0} \right\rangle \\
  &=& \frac{e^{2}}{2mc^{2}} \left\langle|{\bf A}_{1{\rm gc}}|^{2}\right\rangle \;+\; \frac{e}{2c} \left\langle {\bf A}_{1{\rm gc}}\bdot\left\{ {\bf X} + \vb{\rho}_{0},\frac{}{} e\,\wt{\psi}_{1{\rm gc}}\right\}_{0}\right\rangle,
\end{eqnarray*}
so that by combining these two results, we obtain the standard second-order gyrocenter Hamiltonian in the Hamiltonian representation \citep{Brizard_Hahm_2007}:
\[ {\cal H}_{2{\rm gy}}^{\rm Ham} \;=\; \frac{e^{2}}{2mc^{2}} \left\langle|{\bf A}_{1{\rm gc}}|^{2}\right\rangle \;-\; \frac{e}{2} \left\langle \left\{ S_{1},\frac{}{} \psi_{1{\rm gc}}\right\}_{0}\right\rangle, \]
which is expressed in terms of the perturbation electromagnetic potentials only.

\subsubsection{Zero-Larmor-radius limit}

For the purpose of explicit calculations needed for numerical gyrokinetic applications, we now derive the zero-Larmor-radius (ZLR) limit of the second-order gyrocenter Hamiltonian \eqref{eq:H2_gy_final}. Hence, we make the substitutions $({\bf E}_{1{\rm gc}}, {\bf B}_{1{\rm gc}}) \rightarrow ({\bf E}_{1}, {\bf B}_{1})$ in the first term to obtain
\begin{eqnarray*} 
-\frac{1}{2} \left\langle e\,\vb{\rho}_{1{\rm gy}}\bdot\left( {\bf E}_{1{\rm gc}} + \frac{{\bf v}_{0}}{c}\btimes{\bf B}_{1{\rm gc}}\right)\right\rangle_{\rm ZLR} &= & -\,\frac{e}{2}\,\langle\vb{\rho}_{1{\rm gy}}\rangle_{\rm ZLR}\bdot\left( {\bf E}_{1} + \frac{p_{\|}\bhat_{0}}{mc}\btimes{\bf B}_{1}\right) \nonumber \\
 &&-\; \frac{e\Omega_{0}}{2c}\,\left\langle\vb{\rho}_{1{\rm gy}}\btimes\pd{\vb{\rho}_{0}}{\zeta}\right\rangle_{\rm ZLR}\bdot{\bf B}_{1} \\
 & = & \nabla_{\bot}B_{1\|}\bdot\frac{c\mu}{2B_{0}\Omega_{0}}\left( {\bf E}_{1} + \frac{p_{\|}\bhat_{0}}{mc}\btimes{\bf B}_{1}\right) \nonumber \\
 &&+\; \frac{\mu}{2B_{0}}\;|{\bf B}_{1\bot}|^{2}
 \end{eqnarray*}
 The second term yields
 \begin{eqnarray*} 
 \frac{e}{2} \,\left\langle{\cal G}_{1}^{\bf X}\right\rangle_{\rm ZLR}\bdot\left\langle {\bf E}_{1{\rm gc}} \;+\; \frac{{\bf v}_{0}}{c}\btimes{\bf B}_{1{\rm gc}} \right\rangle_{\rm ZLR} &=& \frac{mc^{2}}{2\,B_{0}^{2}} \left|{\bf E}_{1} \;+\; \frac{p_{\|}\bhat_{0}}{mc}\btimes
 {\bf B}_{1}\right|^{2} \\
  &&-\; \nabla_{\bot}B_{1\|}\bdot\frac{c\mu}{2B_{0}\Omega_{0}}\left( {\bf E}_{1} + \frac{p_{\|}\bhat_{0}}{mc}\btimes{\bf B}_{1}\right),
  \end{eqnarray*}
 while the next two terms either vanish or can be omitted as higher-order terms in the ZLR limit. Finally, the last term yields
 \[ -\; \frac{e}{2}\,\left\langle{\cal G}_{1}^{\mu}\right\rangle_{\rm ZLR}\;\pd{\langle\psi_{1{\rm gc}}\rangle_{\rm ZLR}}{\mu} \;=\; \frac{\mu\,B_{1\|}^{2}}{2\,B_{0}}. \]
 Hence, in the ZLR limit, the second-order gyrocenter Hamiltonian \eqref{eq:H2_gy_final} becomes
 \begin{equation}
 {\cal H}_{2{\rm gy}}^{\rm ZLR} \;=\; \frac{mc^{2}}{2\,B_{0}^{2}} \left|{\bf E}_{1} \;+\; \frac{p_{\|}\bhat_{0}}{mc}\btimes {\bf B}_{1}\right|^{2} \;+\; \frac{\mu\,|{\bf B}_{1}|^{2}}{2\,B_{0}} \;\equiv\; K_{2{\rm gy}},
 \label{eq:H2_gy_ZLR}
 \end{equation}
 which is expressed in terms of the perturbed electric and magnetic fields. This second-order gyrocenter Hamiltonian yields the following first-order gyrocenter polarization and magnetization contributions
 \begin{eqnarray}
 \pd{K_{2{\rm gy}}}{{\bf E}_{1}} & = & \frac{mc^{2}}{B_{0}^{2}} \left( {\bf E}_{1} \;+\; \frac{p_{\|}\bhat_{0}}{mc}\btimes {\bf B}_{1}\right) \;=\; \frac{e\bhat_{0}}{\Omega_{0}}\btimes\left( {\bf E}_{1}\btimes\frac{c\bhat_{0}}{B_{0}} \;+\; \frac{p_{\|}}{m}\;\frac{{\bf B}_{1}}{B_{0}} \right), \label{eq:K2_E} \\
 \pd{K_{2{\rm gy}}}{{\bf B}_{1}} & = &  \mu\;\frac{{\bf B}_{1}}{B_{0}} \;+\; \frac{mc^{2}}{B_{0}^{2}} \left( {\bf E}_{1} \;+\; \frac{p_{\|}\bhat_{0}}{mc}\btimes {\bf B}_{1}\right)\btimes\frac{p_{\|}\bhat_{0}}{mc} \nonumber \\
  &=& \mu\;\frac{{\bf B}_{1}}{B_{0}} \;+\; 
 \pd{K_{2{\rm gy}}}{{\bf E}_{1}} \btimes\frac{p_{\|}\bhat_{0}}{mc}. \label{eq:K2_B}
 \end{eqnarray}
 Here, the first-order gyrocenter magnetization kernel \eqref{eq:K2_B} once again is divided into an intrinsic contribution $(\mu{\bf B}_{1}/B_{0})$ and a moving electric-dipole contribution $(\partial K_{2{\rm gy}}/\partial{\bf E}_{1})$.

\subsection{Symplectic Euler-Lagrange and Hamilton gyrocenter equations}

We are now ready to derive explicit gyrocenter equations of motion. First, the gyrocenter Lagrangian is defined as
 \begin{equation}
 L_{\rm gy} \;=\; {\bf P}_{\rm gy}\bdot\dot{\bf X} \;+\; J\,\dot{\zeta} \;-\; H_{\rm gy},
 \label{eq:Lag_gy}
 \end{equation}
 where the gyrocenter canonical momentum is
 \begin{equation}
 {\bf P}_{\rm gy} \;=\; \frac{e}{c} \left( {\bf A}_{0}^{*} \;+\frac{}{} \epsilon\,\langle{\bf A}_{1{\rm gc}}\rangle\right) \;+\; \epsilon\,{\bf P}_{1{\rm gy}}.
 \label{eq:Pgy_symp}
 \end{equation} 
and the gyrocenter Hamiltonian is
 \begin{equation}
H_{\rm gy} \;=\;  \epsilon\;e\langle\Phi_{1{\rm gc}}\rangle \;+\; \frac{p_{\|}^{2}}{2m} + \mu \left( B_{0} \;+\frac{}{} \epsilon\;\langle\langle B_{1\|{\rm gc}}\rangle\rangle \right) \;+\; \epsilon^{2}\,K_{2{\rm gy}}({\bf E}_{1},{\bf B}_{1}),
\label{eq:Hgy_symp}
\end{equation}
where the second-order term is defined in Eq.~\eqref{eq:H2_gy_ZLR}.

The symplectic gyrocenter Euler-Lagrange equations associated with arbitrary variations in $({\bf X}, p_{\|}, J)$ are, respectively,
\begin{eqnarray}
0 & = & e\,{\bf E}_{\rm gy}^{*} \;+\; \frac{e}{c}\dot{\bf X}\btimes{\bf B}_{\rm gy}^{*} \;-\; \dot{p}_{\|}\;{\sf b}_{\rm gy}^{*}, \label{eq:EL_X} \\
0 & = & \dot{\bf X}\bdot{\sf b}_{\rm gy}^{*} \;-\; \partial K_{\rm gy}/\partial p_{\|}, \label{eq:EL_p} \\
0 & = & \dot{\zeta} \;+\; \dot{\bf X}\bdot\partial{\bf P}_{\rm gy}/\partial J \;-\; \partial H_{\rm gy}/\partial J,
\end{eqnarray}
where the effective gyrocenter electric field is
\begin{equation}
e\,{\bf E}_{\rm gy}^{*} \;\equiv\; -\,\nabla H_{\rm gy} - \pd{{\bf P}_{\rm gy}}{t} \;=\; \epsilon\;e\,\left( \langle{\bf E}_{1{\rm gc}}\rangle \;-\; \frac{d_{0}\langle{\bf E}_{1{\rm gc}}\rangle}{dt} \btimes\frac{\bhat_{0}}{\Omega_{0}} \right)  \;-\; \nabla K_{\rm gy},
 \label{eq:Egy_*}
\end{equation}
and the effective gyrocenter magnetic field is
\begin{equation}
{\bf B}_{\rm gy}^{*} \;=\; \nabla\btimes\left(\frac{c}{e}\,{\bf P}_{\rm gy}\right) \;=\; {\bf B}_{0}^{*} + \epsilon\,\langle{\bf B}_{1{\rm gc}}\rangle \;+\;  \epsilon\,\nabla\btimes\left(\frac{c}{e}\,{\bf P}_{1{\rm gy}}\right),
  \label{eq:Bgy_*}
\end{equation}
with ${\bf B}_{0}^{*} \equiv \nabla\btimes{\bf A}_{0}^{*}$ and
\begin{equation}
{\sf b}^{*}_{\rm gy} \;\equiv\; \partial{\bf P}_{\rm gy}/\partial p_{\|} \;=\; \bhat_{0} \;+\; \epsilon\,\langle{\bf B}_{1\bot{\rm gc}}\rangle/B_{0}.
\end{equation}
Here, the perturbed electric field $\langle{\bf E}_{1{\rm gc}}\rangle$ includes its inductive component $-\,c^{-1}\partial_{t}\langle{\bf A}_{1{\rm gc}}\rangle$, and the lowest-order time derivative
\[ \pd{}{t}\left(\langle{\bf E}_{1{\rm gc}}\rangle + \frac{p_{\|}\bhat_{0}}{mc}\btimes\langle{\bf B}_{1{\rm gc}}\rangle\right) \simeq \frac{d_{0}\langle{\bf E}_{1{\rm gc}}\rangle}{dt}  \]
is computed with the help of the gyrokinetic Faraday's Law, with $d_{0}/dt = \partial/\partial t + (p_{\|}/m)\bhat_{0}\bdot\nabla$. We note that the effective gyrocenter electromagnetic fields satisfy the Maxwell equations $\nabla\bdot{\bf B}_{\rm gy}^{*} = 0$ and 
$\partial{\bf B}_{\rm gy}^{*}/\partial t + c\,\nabla\btimes{\bf E}_{\rm gy}^{*} = 0$.

The gyrocenter Euler-Lagrange equations \eqref{eq:EL_X}-\eqref{eq:EL_p} can also be written in Hamiltonian form as
\begin{eqnarray}
\dot{\bf X}  & \equiv & \left\{ {\bf X},\; H_{\rm gy}\right\}_{\rm gy} \;=\; {\bf E}_{\rm gy}^{*}\btimes\frac{c{\sf b}^{*}_{\rm gy}}{B_{\|{\rm gy}}^{**}} + \pd{K_{\rm gy}}{p_{\|}}\,\frac{{\bf B}^{*}_{\rm gy}}{B_{\|{\rm gy}}^{**}}, \label{eq:Xgy_dot} \\
\dot{p}_{\|} & \equiv & \left\{ p_{\|},\; H_{\rm gy}\right\}_{\rm gy} \;=\; e\,{\bf E}_{\rm gy}^{*}\bdot\frac{{\bf B}^{*}_{\rm gy}}{B_{\|{\rm gy}}^{**}}, \label{eq:pgy_dot}
\end{eqnarray}
where $B_{\|{\rm gy}}^{**} \equiv {\sf b}^{*}_{\rm gy}\bdot{\bf B}^{*}_{\rm gy}$. In the expression for the symplectic gyrocenter velocity \eqref{eq:Xgy_dot}, we find the perturbed $E\times B$ velocity (defined in terms of the total magnetic field ${\bf B}_{0} + \epsilon\,\langle{\bf B}_{1{\rm gc}}\rangle$), the polarization drift velocity (involving $d_{0}\langle{\bf E}_{1{\rm gc}}\rangle/dt$), and the total guiding-center drift velocity (i.e., the magnetic gradient and curvature drifts). We note that the identity 
\begin{equation}
\pd{K_{\rm gy}}{p_{\|}}\,\dot{p}_{\|} \;=\; e\,{\bf E}_{\rm gy}^{*}\bdot\pd{K_{\rm gy}}{p_{\|}}\,\frac{{\bf B}^{*}_{\rm gy}}{B_{\|{\rm gy}}^{**}} \;\equiv\; e\,{\bf E}_{\rm gy}^{*}\bdot\dot{\bf X}
\label{eq:xp_dot_id}
\end{equation}
will be useful in our discussion of energy conservation.

\section{\label{sec:symp_VP}Gyrokinetic Variational Principle}

The gyrokinetic Vlasov-Maxwell equations can be derived either from a Low-Lagrange \citep{Sugama_2000}, an Euler \citep{Brizard_PRL_2000,Brizard_PoP_2000,Brizard_2009,Brizard_2010,Brizard_2017b}, a Hamilton-Jacobi \citep{CRP_2004}, or an Euler-Poincar\'{e} \citep{Squire_2013,EH_2020} variational principle.  In recent work, Brizard \& Tronci \cite{Brizard_Tronci_2016} showed how the guiding-center Vlasov-Maxwell equations (derived without a separation between time-independent background and variational dynamical plasma fields) can be explicitly derived according to each of these variational principles. In the present work, the separation of background and perturbed electromagnetic fields introduces a low-frequency gyrokinetic space-time ordering that assumes that the nonuniform background magnetic field is time-independent and non-variational. Applications of Noether's Theorem, which will explicitly take into account the properties of the background magnetic field, follow most naturally from an Eulerian variational principle. In recent work, Hirvijoki {\it et al.} \cite{EH_2020} derived the energy-momentum and angular-momentum conservation laws within an Euler-Poincar\'{e} variational formulation for the Vlasov-Maxwell and drift-kinetic Vlasov-Maxwell equations.

We are now ready to derive the gyrokinetic Vlasov-Maxwell equations from an Eulerian variational principle $\delta{\cal A}_{\rm gy} = 0$, based on the gyrokinetic action functional  \citep{Brizard_PoP_2000}
\begin{equation}
{\cal A}_{\rm gy} \equiv -\,\int {\cal F}_{\rm gy}\,{\cal H}_{\rm gy}\,d^{8}{\cal Z} + \int\frac{d^{4}x}{8\pi} \left( |{\bf E}|^{2} \;-\frac{}{} |{\bf B}|^{2}\right),
\label{eq:A_gy}
\end{equation}
where summation over particle species is implicitly assumed in the first term and the infinitesimal extended phase-space volume element $d^{8}{\cal Z}$ does not include the Jacobian ${\cal J}_{\rm gy}$. Instead, the Jacobian is inserted in the definition of the gyrocenter extended Vlasov density 
\begin{equation}
{\cal F}_{\rm gy} \;\equiv\; {\cal J}_{\rm gy}\,{\cal F} \;\equiv\; {\cal J}_{\rm gy}\,F\,\delta(w - H_{\rm gy}),
\end{equation}
which also includes an energy delta function that enforces the constraint ${\cal H}_{\rm gy} = H_{\rm gy} - w \equiv 0$ in extended gyrocenter phase space.

The variation of the gyrokinetic action functional yields
\begin{equation}
\delta{\cal A}_{\rm gy} \;=\; -\,\int \left(\delta{\cal F}_{\rm gy}\,{\cal H}_{\rm gy} \;+\frac{}{} {\cal F}_{\rm gy}\,\delta{\cal H}_{\rm gy}\right)\,d^{8}{\cal Z} \;+\; \int\frac{d^{4}x}{4\pi} \left( \epsilon\,\delta{\bf E}_{1}\bdot{\bf E} \;-\frac{}{} \epsilon\,
\delta{\bf B}_{1}\bdot{\bf B}\right),
 \label{eq:delta_Lgy}
\end{equation}
where the electromagnetic variations 
\begin{equation}
\left. \begin{array}{rcl}
\delta{\bf E}_{1} &\equiv& -\,\nabla\delta\Phi_{1} \;-\; c^{-1}\partial_{t}\delta {\bf A}_{1} \\
\delta{\bf B}_{1} &\equiv& \nabla\btimes\delta {\bf A}_{1}
\end{array} \right\}
\label{eq:delta_EB}
\end{equation}
satisfy the electromagnetic constraint equations $\nabla\delta{\bf E}_{1} + c^{-1}\partial_{t}\delta{\bf B}_{1} = 0$ and $\nabla\bdot\delta{\bf B}_{1} = 0$, with the background magnetic field ${\bf B}_{0}$ held constant under field variations. The variation of the gyrocenter Hamiltonian \eqref{eq:Psi1_gy}:
\begin{equation}
\delta{\cal H}_{\rm gy} \;=\; \epsilon\,e\;\langle\delta\Phi_{1{\rm gc}}\rangle \;+\; \epsilon\;\mu\,\bhat_{0}\bdot\langle\langle\delta{\bf B}_{1{\rm gc}}\rangle\rangle \;+\; \epsilon^{2} \left(\delta{\bf E}_{1}\bdot\pd{K_{2{\rm gy}}}{{\bf E}_{1}} \;+\;
\delta{\bf B}_{1}\bdot\pd{K_{2{\rm gy}}}{{\bf B}_{1}} \right)
\label{eq:delta_Hgy}
\end{equation}
is expressed in terms of $\delta\Phi_{1}$ and $(\delta{\bf E}_{1},\delta{\bf B}_{1})$, where the second-order terms are calculated in Eqs.~\eqref{eq:K2_E}-\eqref{eq:K2_B} as
\begin{eqnarray}
\pd{K_{2{\rm gy}}}{{\bf E}_{1}} &=& \frac{ec}{B_{0}\Omega_{0}} \left( {\bf E}_{1} \;+\; \frac{p_{\|}\bhat_{0}}{mc}\btimes{\bf B}_{1}\right) = \frac{e\bhat_{0}}{m\Omega_{0}}\btimes{\bf P}_{1}, \nonumber \\
\pd{K_{2{\rm gy}}}{{\bf B}_{1}} &=& \frac{\mu\,{\bf B}_{1}}{B_{0}} \;+\; \pd{K_{2{\rm gy}}}{{\bf E}_{1}}\btimes\frac{p_{\|}\bhat_{0}}{mc}.
\end{eqnarray}
 The variation of the gyrocenter extended Vlasov density $\delta{\cal F}_{\rm gy} \equiv \delta{\cal J}_{\rm gy}\;{\cal F} + {\cal J}_{\rm gy}\;\delta{\cal F}$ is expressed as
\begin{eqnarray}
\delta{\cal F}_{\rm gy} &=& {\cal F} \left( \pd{\delta\vb{\Pi}_{\rm gy}}{p_{\|}}\bdot\frac{e}{c}{\bf B}^{*}_{\rm gy} + {\sf b}^{*}_{\rm gy}\bdot\nabla\btimes\delta\vb{\Pi}_{\rm gy}\right) + {\cal J}_{\rm gy} \left( \{ \delta{\cal S},\; {\cal F}\}_{\rm gy} \;+\frac{}{} 
\delta\vb{\Pi}_{\rm gy}\bdot\{{\bf X},\; {\cal F}\}_{\rm gy} \right) \nonumber \\
 &\equiv& -\;\pd{}{{\cal Z}^{a}}\left( \delta{\cal Z}^{a}\frac{}{}{\cal F}_{\rm gy}\right),
 \label{eq:delta_Fgy}
\end{eqnarray}
where the virtual extended phase-space displacement
\begin{equation}
\delta{\cal Z}^{a} \;\equiv\; \left\{ {\cal Z}^{a},\frac{}{} \delta{\cal S}\right\}_{\rm gy} \;-\; \epsilon\;\delta\vb{\Pi}_{1{\rm gy}}\bdot\left\{ {\bf X},\frac{}{} {\cal Z}^{a}\right\}_{\rm gy}
\end{equation}
is defined in terms of a canonical part generated by $\delta{\cal S}$ and a non-canonical part generated by 
\begin{equation}
\delta\vb{\Pi}_{1{\rm gy}} = \frac{e}{c}\,\langle\delta{\bf A}_{1{\rm gc}}\rangle \;+\; \left(\langle\delta{\bf E}_{1{\rm gc}}\rangle + \frac{p_{\|}\bhat_{0}}{mc}\times\langle\delta{\bf B}_{1{\rm gc}}\rangle\right)\times\frac{e\bhat_{0}}{\Omega_{0}}.
\label{eq:delta_Pi_gy}
\end{equation}

The first two variations in Eq.~\eqref{eq:delta_Lgy} can be combined
\begin{eqnarray}
-\,\delta({\cal F}_{\rm gy}\,{\cal H}_{\rm gy}) &=& -\, {\cal J}_{\rm gy}\{ {\cal F},\; {\cal H}_{\rm gy}\}_{\rm gy}\;\delta{\cal S} \;+\; {\cal F}_{\rm gy}\;\delta L_{\rm gy} \nonumber \\
 &&+\; \pd{}{t}\left({\cal F}_{\rm gy}\,\delta{\cal S}\right) \;+\; \nabla\bdot\left(\dot{\bf X}\;{\cal F}_{\rm gy}\,\delta{\cal S}\right) \;+\; \pd{}{p_{\|}}\left( \dot{p}_{\|}\;{\cal F}_{\rm gy}\,\delta{\cal S}\right),
\label{eq:delta_FH}
\end{eqnarray}
where the variation of the gyrocenter Lagrangian \eqref{eq:Lag_gy}
\begin{eqnarray}
\delta L_{\rm gy} & \equiv & \epsilon\left( \frac{e}{c}\,\langle\delta{\bf A}_{1{\rm gc}}\rangle\bdot\dot{\bf X} \;-\; e\,\langle\delta\Phi_{1{\rm gc}}\rangle \right) \;+\; \epsilon\,\left(\langle\delta{\bf E}_{1{\rm gc}}\rangle \;+\; \frac{p_{\|}\bhat_{0}}{mc}\btimes\langle\delta{\bf B}_{1{\rm gc}}\rangle\right) \bdot\vb{\pi}_{\rm gy} \nonumber \\
 &&-\; \epsilon\;\langle\langle\delta{\bf B}_{1{\rm gc}}\rangle\rangle \bdot \mu\;\bhat_{0}  \;-\; \epsilon^{2} \left(\delta{\bf E}_{1}\bdot\pd{K_{2{\rm gy}}}{{\bf E}_{1}} \;+\; \delta{\bf B}_{1}\bdot\pd{K_{2{\rm gy}}}{{\bf B}_{1}} \right)
 \end{eqnarray}
is expressed in terms of the gyrocenter electric-dipole moment
 \begin{equation}
 \vb{\pi}_{\rm gy} \;\equiv\; (e\bhat_{0}/\Omega_{0})\btimes\dot{\bf X},
 \label{eq:pi_gy}
 \end{equation}
which includes guiding-center \citep{Tronko_Brizard_2015} and gyrocenter \citep{Brizard_2013} contributions. The Lagrangian variation term
 \begin{equation}
 \int_{\cal Z} {\cal F}_{\rm gy}\;\delta L_{1{\rm gy}} \;=\; \int_{x} \left( \frac{1}{c}\,\delta{\bf A}_{1}\bdot{\bf J}_{\rm gy} \;-\; \delta\Phi_{1}\;\varrho_{\rm gy} \;+\; \delta{\bf E}_{1}\bdot\mathbb{P}_{\rm gy} + \delta{\bf B}_{1}\bdot\mathbb{M}_{\rm gy} \right)
\label{eq:Pi_Psi}
\end{equation}
 can be expressed in terms of the gyrocenter charge and current densities
\begin{equation}
\left(\varrho_{\rm gy},\; {\bf J}_{\rm gy}\right) \;\equiv\; \int_{\cal Z} {\cal F}_{\rm gy}\;\langle\delta^{3}({\bf X} + \vb{\rho}_{0} - {\bf x})\rangle\; \left( e\,,\; e\,\dot{\bf X}\right)
\label{eq:rhoJ_gy}
\end{equation}
and the gyrocenter polarization and magnetization
\begin{eqnarray}
\mathbb{P}_{\rm gy} & \equiv & \int_{\cal Z} {\cal F}_{\rm gy} \left( \langle\delta^{3}({\bf X} + \vb{\rho}_{0} - {\bf x})\rangle\;\vb{\pi}_{\rm gy} \;-\; \epsilon\,\delta^{3}\;\pd{K_{2{\rm gy}}}{{\bf E}_{1}} \right), \label{eq:Pol_gy} \\
\mathbb{M}_{\rm gy}  & \equiv & \int_{\cal Z} {\cal F}_{\rm gy}  \left[ -\mu\,\left( \bhat_{0}\,\langle\langle\delta^{3}({\bf X} + \vb{\rho}_{0} - {\bf x})\rangle\rangle \;+\; \epsilon\,\delta^{3}\;\frac{{\bf B}_{1}}{B_{0}} \right) \right. \nonumber \\
 &&\left.+\;  \left(\langle\delta^{3}({\bf X} + \vb{\rho}_{0} - {\bf x})\rangle\;\vb{\pi}_{\rm gy} - \epsilon\,\delta^{3}\;\pd{K_{2{\rm gy}}}{{\bf E}_{1}}\right)\btimes\frac{p_{\|}\bhat_{0}}{mc}\right],
\label{eq:Mag_gy}
\end{eqnarray}
where the delta function $\delta^{3}({\bf X} + \vb{\rho}_{0} - {\bf x}) \equiv \delta^{3}_{\rm gc}$ yields the standard guiding-center finite-Larmor-radius effects (see App.~\ref{sec:Bessel} for additional details) and the first-order corrections due to the second-order gyrocenter Hamiltonian are calculated in the zero-Larmor-radius (ZLR) limit. We note that, in the ZLR limit, the polarization kernel in Eq.~\eqref{eq:Pol_gy} becomes
\begin{eqnarray} 
\vb{\pi}_{1{\rm gy}}^{\rm ZLR} \;-\; \pd{K_{2{\rm gy}}}{{\bf E}_{1}} &=& \frac{e\bhat_{0}}{m\Omega_{0}}\btimes\left({\bf P}_{1} \;+\; \frac{\mu\bhat_{0}}{\Omega_{0}}\btimes\nabla B_{1\|} \right) \;-\; \frac{e\bhat_{0}}{m\Omega_{0}}\btimes{\bf P}_{1} \nonumber \\
 &=& -\;\frac{e\mu}{m\Omega_{0}^{2}}\;\nabla_{\bot}B_{1\|} \;=\; e\,\langle\vb{\rho}_{1{\rm gy}}\rangle_{\rm ZLR},
 \label{eq:pi1_gy_ZLR}
 \end{eqnarray}
which is consistent with the choice made in Eq.~\eqref{eq:rho_1gy} for the first-order gyroangle-averaged gyrocenter displacement. In addition, we note that the gyrocenter magnetization is the sum of the intrinsic magnetic-moment contribution $(-\,\mu\bhat_{0} + \cdots)$ and the moving electric-dipole contribution $(\vb{\pi}_{\rm gy}\btimes p_{\|}\bhat_{0}/mc + \cdots)$. The variation of the Maxwell Lagrangian density can be expressed as 
\begin{eqnarray}
\delta{\bf E}_{1}\bdot{\bf E} - \delta{\bf B}_{1}\bdot{\bf B} &=& \delta {\bf A}_{1}\bdot\left(\frac{1}{c}\pd{\bf E}{t} - \nabla\btimes{\bf B}\right) \;+\; \delta\Phi_{1}\;(\nabla\bdot{\bf E}) \nonumber \\
 &&-\; \pd{}{t}\left(\frac{1}{c}\,\delta {\bf A}_{1}\bdot{\bf E}\right) \;-\; \nabla\bdot\left(\delta\Phi_{1}\,{\bf E} \;+\frac{}{} \delta {\bf A}_{1}\btimes{\bf B}\right).
\label{eq:delta_EB_energy}
\end{eqnarray}

If we now combine Eqs.~\eqref{eq:delta_FH}-\eqref{eq:delta_EB_energy} into the variation of the gyrokinetic action functional \eqref{eq:delta_Lgy}: $\delta{\cal A}_{\rm gy} \equiv \int \delta{\cal L}_{\rm gy}\,d^{4}x$, we obtain the variation of the gyrokinetic Lagrangian density
\begin{eqnarray}
\delta{\cal L}_{\rm gy} & = & -\,\int_{P} {\cal J}_{\rm gy}\{ {\cal F},\; {\cal H}_{\rm gy}\}_{\rm gy}\;\delta{\cal S} \;+\; \frac{\epsilon\,\delta\Phi_{1}}{4\pi} \left( \nabla\bdot\mathbb{D}_{\rm gy} \;-\frac{}{} 4\pi \;\varrho_{\rm gy}  \right) \nonumber \\
 &&+\; \frac{\epsilon}{4\pi}\,\delta {\bf A}_{1}\bdot\left( \frac{1}{c}\pd{\mathbb{D}_{\rm gy}}{t} - \nabla\btimes\mathbb{H}_{\rm gy} + \frac{4\pi}{c}\;{\bf J}_{\rm gy}  \right) \nonumber \\
  &  &+\; \pd{}{t}\left(\int_{P}{\cal F}_{\rm gy}\,\delta{\cal S} - \frac{\epsilon}{4\pi}\,\delta {\bf A}_{1}\bdot\mathbb{D}_{\rm gy}\right) \nonumber \\
  &&+\; \nabla\bdot\left(\int_{P}\dot{\bf X}{\cal F}_{\rm gy}\delta{\cal S} - \frac{\epsilon}{4\pi}\left(\delta\Phi_{1}\mathbb{D}_{\rm gy} + \delta {\bf A}_{1}\times\mathbb{H}_{\rm gy}\right) \right),
  \label{eq:Lag_gy_vp}
\end{eqnarray}
where the gyrocenter macroscopic electromagnetic fields are defined as
\begin{equation}
\left. \begin{array}{rcl}
\mathbb{D}_{\rm gy} & \equiv & \epsilon\,{\bf E}_{1} \;+\; 4\pi\,\mathbb{P}_{\rm gy} \\
\mathbb{H}_{\rm gy} & \equiv & {\bf B}_{0} \;+\; \epsilon\,{\bf B}_{1} \;-\; 4\pi\,\mathbb{M}_{\rm gy}
\end{array} \right\},
\label{eq:DH_def}
\end{equation}
and the variations $(\delta{\cal S}, \delta\Phi_{1}, \delta{\bf A}_{1})$ are assumed to be arbitrary. Variation with respect to $\delta{\cal S}$ yields the gyrokinetic Vlasov equation in extended phase space $\{ {\cal F},\; 
{\cal H}_{\rm gy}\}_{\rm gy} = 0$. If we integrate ${\cal J}_{\rm gy}\{ {\cal F},\; {\cal H}_{\rm gy}\}_{\rm gy}$ over the energy $w$ coordinate, we find
\begin{eqnarray}
0 & = & \int {\cal J}_{\rm gy}\{ {\cal F},\; {\cal H}_{\rm gy}\}_{\rm gy}\;dw \;=\; \int \pd{}{{\cal Z}^{a}}\left({\cal J}_{\rm gy}\,{\cal F}\;\dot{\cal Z}^{a}\right) dw \nonumber \\
 & = & \pd{({\cal J}_{\rm gy}\,F)}{t} + \nabla\bdot\left({\cal J}_{\rm gy}\,F\frac{}{}\dot{\bf X}\right) + \pd{}{p_{\|}}\left({\cal J}_{\rm gy}\,F\frac{}{}\dot{p}_{\|}\right) \nonumber \\
  &\equiv& {\cal J}_{\rm gy} \left( \pd{F}{t} + \dot{\bf X}\bdot\nabla F + \dot{p}_{\|}\;\pd{F}{p_{\|}}\right),
  \label{eq:Vlasov_gy}
\end{eqnarray}
where we have used the Liouville theorem \eqref{eq:Liouville} to obtain the last expression in order to recover the gyrokinetic Vlasov equation.

Next, the variation with respect to the electromagnetic potentials $(\delta\Phi_{1}, \delta{\bf A}_{1})$ yield the macroscopic gyrokinetic Maxwell equations
\begin{eqnarray}
\nabla\bdot\mathbb{D}_{\rm gy} & = & 4\pi\;\varrho_{\rm gy}, \label{eq:div_D} \\
\nabla\btimes\mathbb{H}_{\rm gy} & = & \frac{1}{c}\pd{\mathbb{D}_{\rm gy}}{t} +\; \frac{4\pi}{c} \;{\bf J}_{\rm gy}, \label{eq:curl_H} 
 \end{eqnarray}
 which can also be expressed as the microscopic Maxwell equations
\begin{eqnarray}
\nabla\bdot\epsilon\,{\bf E}_{1} & = & 4\pi\,\left(\varrho_{\rm gy} \;-\frac{}{} \nabla\bdot\mathbb{P}_{\rm gy}\right), \label{eq:div_E} \\
\nabla\btimes\left({\bf B}_{0} \;+\frac{}{} \epsilon\,{\bf B}_{1}\right) & = & \frac{\epsilon}{c}\pd{{\bf E}_{1}}{t} + \frac{4\pi}{c} \left({\bf J}_{\rm gy} + \pd{\mathbb{P}_{\rm gy}}{t} + c\,\nabla\times\mathbb{M}_{\rm gy}\right). \label{eq:curl_B} 
 \end{eqnarray}
These equations are complemented by Faraday's Law
 \begin{equation}
 \pd{{\bf B}_{1}}{t} \;+\; c\,\nabla\btimes{\bf E}_{1} \;=\; 0
 \label{eq:Faraday}
 \end{equation}
 and $\nabla\bdot{\bf B}_{1} = 0$. Now that the gyrokinetic Vlasov-Maxwell equations \eqref{eq:Vlasov_gy}-\eqref{eq:curl_H} have been derived from a variational principle, we now use the remaining part of the gyrokinetic Lagrangian density variation \eqref{eq:Lag_gy_vp} to derive exact conservation laws. 
 
\section{\label{sec:symp_laws}Symplectic Gyrokinetic Conservation Laws}

The variational derivation of the reduced Vlasov-Maxwell equations guarantees that these reduced equations satisfy exact energy-momentum conservation laws \citep{Pfirsch_Morrison_1985,CRP_2004,Brizard_2008}. In particular, the exact conservation of the gyrokinetic Vlasov-Maxwell energy \citep{Brizard_1989,Brizard_2010} has played an important role in the numerical implementation of the energy-conserving gyrokinetic equations \citep{Garbet_2010}. The gyrokinetic angular-momentum conservation law (derived consistently with a variational principle) has so far only been discussed in the case of electrostatic potential fluctuations \citep{Scott_Smirnov_2010,Brizard_Tronko_2011}, while the case of full electromagnetic fluctuations was discussed recently by Hirvijoki {\it et al.} \cite{EH_2020} in the drift-kinetic limit. In the present Section, we derive the gyrokinetic Noether energy-momentum equations and extract exact energy-momentum and angular-momentum conservation laws for the gyrokinetic Vlasov-Maxwell equations \eqref{eq:Vlasov_gy}-\eqref{eq:curl_H}.

For this purpose, the remaining terms in Eq.~\eqref{eq:Lag_gy_vp} are combined to yield the gyrokinetic Noether equation
\begin{eqnarray}
\delta{\cal L}_{\rm gy} &=& \pd{}{t}\left(\int_{P}{\cal F}_{\rm gy}\,\delta{\cal S} - \frac{\epsilon}{4\pi c}\,\delta {\bf A}_{1}\bdot\mathbb{D}_{\rm gy}\right) \nonumber \\
 &&+\; \nabla\vb{\cdot}\left[\int_{P}\dot{\bf X}{\cal F}_{\rm gy}\delta{\cal S} - \frac{\epsilon}{4\pi}\left(\delta\Phi_{1}\mathbb{D}_{\rm gy} \;+\frac{}{} \delta {\bf A}_{1}\times\mathbb{H}_{\rm gy}\right) \right],
\label{eq:Noether_1} 
\end{eqnarray}
where the variations are now explicitly expressed in terms of the space-time displacements $\delta{\bf x}$ and $\delta t$:
\begin{equation}
\left. \begin{array}{rcl}
\delta{\cal S} & \equiv & {\bf P}_{\rm gy}\bdot\delta{\bf x} \;-\; w\,\delta t \\
\delta\Phi_{1} & \equiv & -\,\delta{\bf x}\bdot\nabla\Phi_{1} - \delta t\,\partial\Phi_{1}/\partial t \;=\; {\bf E}_{1}\bdot\delta{\bf x} \;-\; c^{-1}\partial\delta\chi_{1}/\partial t \\
\delta{\bf A}_{1} & \equiv & -\,\delta{\bf x}\bdot\nabla{\bf A}_{1} - \delta t\,\partial{\bf A}_{1}/\partial t \;=\;  {\bf E}_{1}\,c\,\delta t \;+\; \delta{\bf x}\btimes{\bf B}_{1} \;+\; \nabla\delta\chi_{1} 
\end{array} \right\},
\label{eq:delta_SphiA}
\end{equation}
with the gauge variation $\delta\chi_{1}$ defined as $\delta\chi_{1} \equiv \Phi_{1}\,c\,\delta t \;-\; {\bf A}_{1}\bdot\delta{\bf x}$. Upon rearranging the gauge variation $\delta\chi_{1}$, and using the identity
\begin{eqnarray*} 
 &&-\,\pd{}{t}\left(\nabla\delta\chi_{1}\bdot\frac{}{}\mathbb{D}_{\rm gy}\right) + \nabla\bdot\left(\pd{\delta\chi_{1}}{t}\;\mathbb{D}_{\rm gy} - c\,\nabla\delta\chi_{1}\btimes\mathbb{H}_{\rm gy}\right)  \\
 &&\;\;=\;  \pd{}{t}\left(\delta\chi_{1}\frac{}{}\nabla\bdot\mathbb{D}_{\rm gy}\right) \;-\; \nabla\bdot\left[ \delta\chi_{1}\left( \pd{\mathbb{D}_{\rm gy}}{t} - c\,\nabla\btimes\mathbb{H}_{\rm gy}\right) \right],
 \end{eqnarray*}
with the macroscopic gyrokinetic Maxwell equations \eqref{eq:div_D}-\eqref{eq:curl_H}, we obtain the gauge-invariant form of the gyrokinetic Noether equation \eqref{eq:Noether_1}:
\begin{equation} 
\delta{\cal L}_{\rm gy} \;=\; \partial\delta{\cal N}_{\rm gy}/\partial t + \nabla\bdot\delta\vb{\Gamma}_{\rm gy},
\label{eq:Noether_gy}
\end{equation}
where the action-density variation is
\begin{equation}
\delta{\cal N}_{\rm gy} \;=\; \int_{P}{\cal F}_{\rm gy}\,\left( \delta{\cal S} + \epsilon\,\frac{e}{c}\langle\delta\chi_{1{\rm gc}}\rangle\right) \;-\; \left(\epsilon\,{\bf E}_{1}\,\delta t + \delta{\bf x}\btimes\frac{\epsilon}{c}\,{\bf B}_{1}\right) \bdot\frac{\mathbb{D}_{\rm gy}}{4\pi} ,
 \end{equation}
and the action-density-flux variation is
\begin{eqnarray}
\delta\vb{\Gamma}_{\rm gy} &=& \int_{P}\dot{\bf X}\;{\cal F}_{\rm gy}\left( \delta{\cal S} + \epsilon\,\frac{e}{c}\langle\delta\chi_{1{\rm gc}}\rangle\right) \;-\; \delta{\bf x}\bdot\left( \frac{\epsilon}{4\pi}{\bf E}_{1}\mathbb{D}_{\rm gy} \right) \nonumber \\
 &&+\; \frac{\epsilon}{4\pi}\left({\bf E}_{1}\,c\,\delta t \;+\frac{}{} \delta{\bf x}\btimes{\bf B}_{1}\right)\times\mathbb{H}_{\rm gy}.
\end{eqnarray}
Here, the gauge-invariant terms are
\begin{equation}
\delta{\cal S} + \epsilon\,\frac{e}{c}\langle\delta\chi_{1{\rm gc}}\rangle \;=\; \left( {\bf P}_{\rm gy} - \epsilon\frac{e}{c}\,\langle{\bf A}_{1{\rm gc}}\rangle\right)\bdot\delta{\bf x} \;-\; \left( w \;-\frac{}{} \epsilon\,e\,\langle\Phi_{1{\rm gc}}\rangle\right)\,\delta t,
\end{equation}
with
\begin{eqnarray*}
{\bf P}_{\rm gy} - \frac{e}{c}\,\epsilon\,\langle{\bf A}_{1{\rm gc}}\rangle & = &  \frac{e}{c}\,{\bf A}_{0}^{*} +  \epsilon \left( \langle{\bf E}_{1{\rm gc}}\rangle \;+\; 
\frac{p_{\|}\bhat_{0}}{mc}\btimes\langle{\bf B}_{1{\rm gc}}\rangle\right)\btimes\frac{e\bhat_{0}}{\Omega_{0}} \;\equiv\;
\frac{e}{c}\,{\bf A}_{0}^{*}  \;+\; \epsilon\,{\bf P}_{1{\rm gy}}, \\
 w - e\,\epsilon\,\langle\Phi_{1{\rm gc}}\rangle & = &  (w - H_{\rm gy}) \;+\; K_{\rm gy}.
 \end{eqnarray*}
We note that the guiding-center vector potential ${\bf A}_{0}^{*}$, which yields the unperturbed background magnetic field ${\bf B}_{0}^{*} = \nabla\btimes{\bf A}_{0}^{*}$, is not subject to a gauge transformation.
 
A complete expression for the gyrokinetic Noether equation \eqref{eq:Noether_gy} also requires an explicit expression for the Lagrangian variation $\delta{\cal L}_{\rm gy}$ on the left side of Eq.~\eqref{eq:Noether_gy}. For the derivation of the momentum-energy conservation laws, we consider the specific space-time variations of the gyrokinetic Lagrangian density
\begin{eqnarray}
\delta{\cal L}_{\rm gy} &=& -\,\left(\delta t \pd{}{t} + \delta{\bf x}\bdot\nabla\right)\left[ \frac{1}{8\pi} \left( \epsilon^{2}|{\bf E}_{1}|^{2} \;-\frac{}{} |{\bf B}|^{2}\right) \right]  \nonumber \\
 &&-\; \delta{\bf x}\bdot\left[ \nabla{\bf B}_{0}\bdot\frac{\bf B}{4\pi} \;+\; \int_{\bf P} {\cal J}_{\rm gy}\,F \left(\nabla^{\prime}K_{\rm gy} - \nabla^{\prime}{\bf P}_{\rm gy}\bdot\dot{\bf X}\right) \right],
 \label{eq:delta_Lgy_Noether}
\end{eqnarray}
where the gradient operator $\nabla^{\prime}$ only takes into account the non-uniformity of the background magnetic field, i.e., the first-order fields $\langle{\bf E}_{1{\rm gc}}\rangle$ and $\langle{\bf B}_{1{\rm gc}}\rangle$ are frozen at a fixed position 
${\bf x} = {\bf X} + \vb{\rho}_{0}$ so that, for example, $\nabla^{\prime}\langle\langle B_{1\|{\rm gc}}\rangle\rangle = \nabla\bhat_{0}\bdot\langle\langle{\bf B}_{1{\rm gc}}\rangle\rangle$. In addition, the $w$-integration was performed to leave the standard gyrocenter Vlasov distribution $F({\bf X},p_{\|},\mu,t)$, with $\int_{\bf P}$ now denoting an integration over $(p_{\|},\mu)$.

The final form of the gyrokinetic Noether equation is obtained by equating Eqs.~\eqref{eq:Noether_gy} and \eqref{eq:delta_Lgy_Noether}, where the virtual space-time displacements $(\delta{\bf x}, \delta t)$ appear explicitly. This form of the Noether Theorem relies on the constrained variations \eqref{eq:delta_EB}, \eqref{eq:delta_Fgy}, and \eqref{eq:delta_SphiA}, which is in contrast to the more traditional formulation based on the connection between conservation laws and symmetries of the Vlasov-Maxwell Lagrangian (see, for example, Ref.~\cite{EH_2020} and references therein). Here, for each conservation law derived from our gyrokinetic Noether equation, we also present an explicit proof based on the gyrokinetic Vlasov-Maxwell equations \eqref{eq:Vlasov_gy}-\eqref{eq:curl_H}.

\subsection{Gyrokinetic energy conservation law}

Since the background magnetic field ${\bf B}_{0}$ is time-independent, the total energy associated with the gyrokinetic Vlasov-Maxwell equations \eqref{eq:Vlasov_gy}-\eqref{eq:curl_H} is conserved. We derive the energy conservation law from the gyrokinetic Noether equation by setting $\delta t \neq 0$ and $\delta{\bf x} = 0$ in Eqs.~\eqref{eq:Noether_gy}  and \eqref{eq:delta_Lgy_Noether}, which yields the gyrokinetic energy conservation law 
\begin{equation}
\partial{\cal E}_{\rm gy}/\partial t \;+\; \nabla\bdot{\bf S}_{\rm gy} \;=\; 0, 
\label{eq:energy_gy}
\end{equation}
where the gyrokinetic energy density is
\begin{eqnarray}
{\cal E}_{\rm gy} & = & \int_{\bf P} {\cal J}_{\rm gy}\,F\;K_{\rm gy} + \epsilon\,{\bf E}_{1}\bdot\mathbb{P}_{\rm gy} + \frac{1}{8\pi}\,\left( \epsilon^{2}|{\bf E}_{1}|^{2} \;+\frac{}{} |{\bf B}|^{2} \right) \nonumber \\
 & = &  \int_{\bf P} {\cal J}_{\rm gy}\,F\left[ \frac{p_{\|}^{2}}{2m} \;+\; \mu\,\left( B_{0} \;+\frac{}{} \epsilon\,\langle\langle B_{1\|{\rm gc}}\rangle\rangle \right) \;+\; \epsilon\,\langle{\bf E}_{1{\rm gc}}\rangle\bdot\left(\frac{e\bhat_{0}}{\Omega_{0}}\btimes\dot{\bf X}\right)  \right. \nonumber \\
  &&\left.+\; \epsilon^{2} \left(K_{2{\rm gy}} \;-\; {\bf E}_{1}\bdot\pd{K_{2{\rm gy}}}{{\bf E}_{1}}\right)\right]  \;+\; \frac{1}{8\pi}\,\left( \epsilon^{2}|{\bf E}_{1}|^{2} \;+\frac{}{} |{\bf B}|^{2} \right),
 \label{eq:E_density}
\end{eqnarray}
while the gyrokinetic energy-density flux is
\begin{equation}
{\bf S}_{\rm gy} \;=\; \int_{\bf P} {\cal J}_{\rm gy}\,F\,K_{\rm gy}\;\dot{\bf X} + \frac{c}{4\pi}\,\epsilon\,{\bf E}_{1}\btimes\mathbb{H}_{\rm gy},
 \label{eq:S_density}
 \end{equation}
 where the polarization and magnetization $(\mathbb{P}_{\rm gy}, \mathbb{M}_{\rm gy})$ are defined in Eqs.~\eqref{eq:Pol_gy}-\eqref{eq:Mag_gy}, with $\mathbb{H}_{\rm gy}$ defined in Eq.~\eqref{eq:DH_def}. In addition, we note that the gyrokinetic polarization  
and magnetization $(\mathbb{P}_{\rm gy}, \mathbb{M}_{\rm gy})$ include the full gyrocenter velocity $\dot{\bf X}$ defined in Eq.~\eqref{eq:Xgy_dot}, which is expressed in terms of the effective electric and magnetic fields \eqref{eq:Egy_*}-\eqref{eq:Bgy_*}. We also note that, as shown by Burby {\it et al.} \cite{Burby_2015}, the gyrokinetic Vlasov-Maxwell Hamiltonian functional is naturally derived from the gyrokinetic energy density \eqref{eq:E_density}.
 
 The explicit proof of energy conservation proceeds as follows. First, we begin with
 \begin{eqnarray}
 \pd{{\cal E}_{\rm gy}}{t} &=& \int_{\bf P} \left[ \pd{({\cal J}_{\rm gy}\,F)}{t}\;K_{\rm gy} \;+\; {\cal J}_{\rm gy}\,F\;\left( \pd{{\bf E}_{1}}{t}\bdot\pd{K_{\rm gy}}{{\bf E}_{1}} + \pd{{\bf B}_{1}}{t}\bdot\pd{K_{\rm gy}}{{\bf B}_{1}} \right) \right] \nonumber \\
  &&+\; \frac{\epsilon\,{\bf E}_{1}}{4\pi}\bdot\pd{\mathbb{D}_{\rm gy}}{t} \;+\; \epsilon\,\pd{{\bf E}_{1}}{t}\bdot\mathbb{P}_{\rm gy} \;+\; \frac{\bf B}{4\pi}\bdot\epsilon\,\pd{{\bf B}_{1}}{t}.
 \label{eq:Eproof_1}
  \end{eqnarray}
 Using the phase-space divergence form \eqref{eq:Vlasov_gy} of the gyrokinetic Vlasov equation, the first term on the right can be expressed as
 \begin{equation}
 \int_{\bf P} \pd{({\cal J}_{\rm gy}\,F)}{t}\;K_{\rm gy} \;=\; -\;\nabla\bdot\left(\int_{\bf P}{\cal J}_{\rm gy}\,F\,K_{\rm gy}\;\dot{\bf X}\right) \;+\; \int_{\bf P}{\cal J}_{\rm gy}\,F \left( \pd{K_{\rm gy}}{p_{\|}}\;\dot{p}_{\|} \;+\; \dot{\bf X}\bdot\nabla K_{\rm gy}\right),
  \end{equation}
 while, using the definitions \eqref{eq:Pol_gy}-\eqref{eq:Mag_gy} of the gyrokinetic polarization and magnetization, the gyrokinetic kinetic terms in Eq.~\eqref{eq:Eproof_1} can be expressed
 \begin{eqnarray}
 \int_{\bf P} {\cal J}_{\rm gy}\,F\;\pd{{\bf E}_{1}}{t}\vb{\cdot}\pd{K_{\rm gy}}{{\bf E}_{1}} & = & -\,\epsilon\,\pd{{\bf E}_{1}}{t}\vb{\cdot}\mathbb{P}_{\rm gy} + \int_{\bf P} {\cal J}_{\rm gy}\,F\;\vb{\pi}_{\rm gy}\vb{\cdot} \pd{}{t}\left(\epsilon\frac{}{}
 \langle{\bf E}_{1{\rm gc}}\rangle\right),  \\
\int_{\bf P} {\cal J}_{\rm gy}\,F\;\pd{{\bf B}_{1}}{t}\vb{\cdot}\pd{K_{\rm gy}}{{\bf B}_{1}} & = & -\,\epsilon\,\pd{{\bf B}_{1}}{t}\vb{\cdot}\mathbb{M}_{\rm gy} + \int_{\bf P} {\cal J}_{\rm gy}\,F\;\vb{\pi}_{\rm gy}\vb{\cdot}\pd{}{t} \left(\frac{p_{\|}\bhat_{0}}{mc}\vb{\times}\epsilon\,\langle{\bf B}_{1{\rm gc}}\rangle\right).
\end{eqnarray}
By combining these expressions, Eq.~\eqref{eq:Eproof_1} becomes
\begin{eqnarray}
 \pd{{\cal E}_{\rm gy}}{t} & = & -\;\nabla\bdot\left(\int_{\bf P}{\cal J}_{\rm gy}\,F\,K_{\rm gy}\;\dot{\bf X}\right) + \frac{\mathbb{H}_{\rm gy}}{4\pi}\bdot\epsilon\,\pd{{\bf B}_{1}}{t} + \frac{\epsilon\,{\bf E}_{1}}{4\pi}\bdot\pd{\mathbb{D}_{\rm gy}}{t} 
   \label{eq:Eproof_2} \\
  & &+ \int_{\bf P}{\cal J}_{\rm gy}\,F\left[ \vb{\pi}_{\rm gy}\bdot\pd{}{t}\left( \epsilon\,\langle{\bf E}_{1{\rm gc}}\rangle + \frac{p_{\|}\bhat_{0}}{mc}\btimes\epsilon\,\langle{\bf B}_{1{\rm gc}}\rangle\right) + \pd{K_{\rm gy}}{p_{\|}}\;\dot{p}_{\|} + 
  \dot{\bf X}\bdot\nabla K_{\rm gy}\right],
\nonumber
\end{eqnarray}
where we introduced the definition of the polarization $\mathbb{P}_{\rm gy}$ and the definitions \eqref{eq:DH_def} for the macroscopic fields $(\mathbb{D}_{\rm gy}, \mathbb{H}_{\rm gy})$. Next, we use Faraday's Law \eqref{eq:Faraday} to write
\[ \frac{\mathbb{H}_{\rm gy}}{4\pi}\bdot\epsilon\,\pd{{\bf B}_{1}}{t}  \;=\; -\;\frac{c\,\mathbb{H}_{\rm gy}}{4\pi}\bdot\nabla\btimes\epsilon\,{\bf E}_{1} \;=\; -\;\nabla\bdot\left( \frac{c}{4\pi}\,\epsilon\,{\bf E}_{1}\btimes\mathbb{H}_{\rm gy}\right) \;-\; \frac{\epsilon\,{\bf E}_{1}}{4\pi}\bdot c\,\nabla\btimes\mathbb{H}_{\rm gy}, \]
so that Eq.~\eqref{eq:Eproof_2} becomes
\begin{eqnarray}
\pd{{\cal E}_{\rm gy}}{t} + \nabla\bdot{\bf S}_{\rm gy} &=& -\;\frac{\epsilon\,{\bf E}_{1}}{4\pi}\bdot\left( c\,\nabla\btimes\mathbb{H}_{\rm gy} \;-\; \pd{\mathbb{D}_{\rm gy}}{t} \right) \nonumber \\
 &&+\; \int_{\bf P}{\cal J}_{\rm gy}\,F\left[ \dot{\bf X}\bdot\epsilon\,\pd{{\bf P}_{1{\rm gy}}}{t} \;+\; \pd{K_{\rm gy}}{p_{\|}}\;\dot{p}_{\|} \;+\;   \dot{\bf X}\bdot\nabla K_{\rm gy} \right],
 \label{eq:Eproof_3}
 \end{eqnarray}
where we reconstructed the gyrokinetic energy-density flux \eqref{eq:S_density} on the left side of Eq.~\eqref{eq:Eproof_3}. Lastly, we use the macroscopic gyrokinetic Maxwell equation \eqref{eq:curl_H} to obtain
\begin{equation}
\pd{{\cal E}_{\rm gy}}{t} + \nabla\bdot{\bf S}_{\rm gy} \;=\;  \int_{\bf P}{\cal J}_{\rm gy}\,F\left( \pd{K_{\rm gy}}{p_{\|}}\;\dot{p}_{\|} \;-\; \dot{\bf X}\bdot\,e\,{\bf E}_{\rm gy}^{*} \right),
\label{eq:E_proof4}
 \end{equation}
 where we introduced the definition \eqref{eq:Egy_*} of the effective gyrocenter electric field ${\bf E}_{\rm gy}^{*}$. Using the identity \eqref{eq:xp_dot_id}, the right side of Eq.~\eqref{eq:E_proof4} is shown to vanish and we readily recover the exact gyrokinetic energy conservation law. 
 
 Lastly, we note that the proof of the gyrokinetic energy conservation law \eqref{eq:E_proof4} is exact even when the second-order gyrokinetic Hamiltonian $K_{2{\rm gy}}$ is removed from our gyrokinetic Vlasov-Maxwelll model, i.e., when the gyrocenter kinetic energy is truncated at first order: $K_{\rm gy} = p_{\|}^{2}/2m + \mu\, (B + \epsilon\,\langle\langle B_{1\|{\rm gc}}\rangle\rangle)$.
 
\subsection{Gyrokinetic Noether momentum equation}

Because the background magnetic field ${\bf B}_{0}$ considered in standard gyrokinetic Vlasov-Maxwell theory is weakly non-uniform (i.e., it serves to magnetically confine charged particles in accordance with the guiding-center approximation), a general gyrokinetic Vlasov-Maxwell momentum conservation law does not exist. Indeed, according to Noether's Theorem, momentum is conserved only in directions corresponding to symmetries of the background magnetic field. Before we derive the gyrokinetic angular-momentum conservation law associated with an axisymmetric background magnetic field, we wish to show that the gyrokinetic Noether momentum equation, from which our exact angular-momentum conservation law will be derived, is consistent with the gyrokinetic Vlasov-Maxwell equations \eqref{eq:Vlasov_gy}-\eqref{eq:curl_H}.

We begin with the gyrokinetic Noether momentum equation derived by setting $\delta t = 0$ and $\delta{\bf x} \neq 0$ in Eqs.~\eqref{eq:Noether_gy}  and \eqref{eq:delta_Lgy_Noether}:
\begin{equation}
 \pd{\vb{\cal P}^{*}_{\rm gy}}{t} + \nabla\bdot{\sf T}^{*}_{\rm gy} =  \int_{\bf P} {\cal J}_{\rm gy}\,F\left( \frac{e}{c}\nabla{\bf A}_{0}^{*}\bdot\dot{\bf X} + \epsilon\,\nabla^{\prime}{\bf P}_{1{\rm gy}}\bdot\dot{\bf X} - \nabla^{\prime}K_{\rm gy}\right)  - 
 \nabla{\bf B}_{0}\bdot\frac{\bf B}{4\pi},
 \label{eq:momentum}
 \end{equation}
where the gyrokinetic  canonical momentum density is defined as
\begin{equation} 
\vb{\cal P}^{*}_{\rm gy} \;=\; \int_{\bf P} {\cal J}_{\rm gy}\,F\,\left( \frac{e}{c}\,{\bf A}_{0}^{*} \;+\; \epsilon\,{\bf P}_{1{\rm gy}}\right) \;+\; \frac{\mathbb{D}_{\rm gy}}{4\pi c}\btimes\epsilon\,{\bf B}_{1} 
\label{eq:Pgy_can}
\end{equation}
and the gyrokinetic  canonical stress tensor is defined as
\begin{eqnarray}
{\sf T}^{*}_{\rm gy} & = &  \int_{\bf P} {\cal J}_{\rm gy}\,F\;\dot{\bf X}\,\left( \frac{e}{c}\,{\bf A}_{0}^{*} \;+\; \epsilon\,{\bf P}_{1{\rm gy}}\right) \;-\; \frac{\epsilon}{4\pi} \left( \mathbb{D}_{\rm gy}\,{\bf E}_{1} \;+\frac{}{} {\bf B}_{1}\,\mathbb{H}_{\rm gy}\right) \nonumber \\
 &  &+\; \mathbb{I} \left[ \frac{1}{8\pi} \left( \epsilon^{2}\,|{\bf E}_{1}|^{2} \;-\frac{}{} |{\bf B}|^{2} \right) + \frac{\epsilon}{4\pi}\,{\bf B}_{1}\bdot\mathbb{H}_{\rm gy}\right],
\label{eq:Tgy_can}
\end{eqnarray}
where $ \mathbb{I}$ denotes the identity matrix. We will return to the gyrokinetic Noether canonical  momentum equation \eqref{eq:momentum} when we derive the gyrokinetic canonical angular-momentum conservation law. We note that, while the gyrokinetic stress tensor \eqref{eq:Tgy_can} is manifestly not symmetric, the exact conservation of the gyrokinetic angular-momentum will follow from the right side of Eq.~\eqref{eq:Pphi} vanishing exactly through an intricate series of cancellations.

\subsubsection{Perturbed gyrokinetic Noether momentum equation}

 We first would like to show that Eq.~\eqref{eq:momentum} is an exact consequence of the gyrokinetic Vlasov-Maxwell equations \eqref{eq:Vlasov_gy}-\eqref{eq:curl_H}. We begin with simplifying the gyrokinetic Noether canonical momentum equation \eqref{eq:momentum} by using the phase-space divergence form \eqref{eq:Vlasov_gy} of the gyrokinetic Vlasov equation to obtain 
\[ \pd{}{t}\left( \int_{\bf P} {\cal J}_{\rm gy}\,F\;\frac{e}{c}\,{\bf A}_{0}^{*}\right) \;=\; -\;\nabla\bdot \left(\int_{\bf P} {\cal J}_{\rm gy}\,F\;\dot{\bf X}\frac{e}{c}\,{\bf A}_{0}^{*}\right) \;+\;  \int_{\bf P} {\cal J}_{\rm gy}\,F\; \left( \frac{e}{c}\;\dot{\bf X}\bdot\nabla{\bf A}_{0}^{*} \;+\; \dot{p}_{\|}\;\bhat_{0}\right), \]
which allows us to obtain the perturbed gyrokinetic Noether momentum equation
\begin{eqnarray}
\pd{\vb{\cal P}_{1{\rm gy}}}{t} \;+\; \nabla\bdot{\sf T}_{1{\rm gy}} &=& \epsilon^{-1}\int_{\bf P} {\cal J}_{\rm gy}\,F\left( \frac{e}{c}\dot{\bf X}\btimes{\bf B}_{0}^{*} \;+\; \epsilon\,\nabla^{\prime}{\bf P}_{1{\rm gy}}\bdot\dot{\bf X} \;-\; \nabla^{\prime}K_{\rm gy} - \dot{p}_{\|}\,\bhat_{0} \right) \nonumber \\
 &&-\; \nabla{\bf B}_{0}\bdot\frac{{\bf B}_{1}}{4\pi},
\label{eq:momentum_first}
\end{eqnarray}
where the perturbed gyrokinetic momentum density is defined as
\begin{equation} 
\vb{\cal P}_{1{\rm gy}} \;=\; \int_{\bf P} {\cal J}_{\rm gy}\,F\,{\bf P}_{1{\rm gy}} \;+\; \frac{\mathbb{D}_{\rm gy}}{4\pi c}\btimes{\bf B}_{1},
\end{equation}
and the perturbed gyrokinetic stress tensor is defined as
\begin{eqnarray} 
{\sf T}_{1{\rm gy}} &=& \int_{\bf P} {\cal J}_{\rm gy}\,F\;\dot{\bf X}\,{\bf P}_{1{\rm gy}} \;-\; \frac{1}{4\pi} \left( \mathbb{D}_{\rm gy}\,{\bf E}_{1} \;+\frac{}{} {\bf B}_{1}\,\mathbb{H}_{\rm gy}\right) \nonumber \\
 &&+\; \mathbb{I} \left[ \frac{\epsilon}{8\pi} \left( |{\bf E}_{1}|^{2} \;+\frac{}{} |{\bf B}_{1}|^{2} \right) \;-\; {\bf B}_{1}\bdot\mathbb{M}_{\rm gy}\right].
\label{eq:stress_first}
\end{eqnarray}
We note that the first term on the right side of Eq.~\eqref{eq:momentum_first} includes the unperturbed form of the Euler-Lagrange equation \eqref{eq:EL_X}: $(e/c)\,\dot{\bf X}_{0}\btimes{\bf B}_{0}^{*} - \nabla K_{\rm gc} - \dot{p}_{0\|}\,\bhat_{0} = 0$. Hence, in the absence of electromagnetic-field  perturbations $(\epsilon = 0)$, the perturbed gyrokinetic Noether momentum equation \eqref{eq:momentum_first} is identically satisfied.

The proof that the perturbed gyrokinetic Noether momentum equation \eqref{eq:momentum_first} follows from the gyrokinetic Vlasov-Maxwell equations \eqref{eq:Vlasov_gy}-\eqref{eq:curl_H} resumes by evaluating the partial time derivative
\begin{eqnarray}
\pd{\vb{\cal P}_{1{\rm gy}}}{t} & = & -\;\int_{\bf P} {\bf P}_{1{\rm gy}} \left[ \nabla\bdot\left({\cal J}_{\rm gy}\,F\;\dot{\bf X}\right) \;+\; \pd{}{p_{\|}}\left({\cal J}_{\rm gy}\,F\;\dot{p}_{\|}\right) \right] \;+\; \int_{\bf P} {\cal J}_{\rm gy}\,F\,\pd{{\bf P}_{1{\rm gy}}}{t} \nonumber \\
 & &+\; \left(\nabla\btimes\mathbb{H}_{\rm gy} \;-\; \frac{4\pi}{c}\,{\bf J}_{\rm gy}\right)\btimes\frac{{\bf B}_{1}}{4\pi} \;-\; \frac{\mathbb{D}_{\rm gy}}{4\pi}\btimes(\nabla\btimes{\bf E}_{1}),
 \end{eqnarray}
where we used Eq.~\eqref{eq:Vlasov_gy}, which yields
\begin{eqnarray}
\pd{\vb{\cal P}_{1{\rm gy}}}{t} \;+\; \nabla\bdot{\sf T}_{1{\rm gy}}   & = & \int_{\bf P} {\cal J}_{\rm gy}\,F\, \left( \pd{{\bf P}_{1{\rm gy}}}{t} + \dot{\bf X}\bdot\nabla{\bf P}_{1{\rm gy}} + \dot{p}_{\|}\;\pd{{\bf P}_{1{\rm gy}}}{p_{\|}} \right) \;-\; \nabla{\bf B}_{0}\bdot
\frac{{\bf B}_{1}}{4\pi} \nonumber \\
 &&- \left( \varrho_{\rm gy}\,{\bf E}_{1} + \frac{1}{c}\,{\bf J}_{\rm gy}\btimes{\bf B}_{1}\right) - \left( \nabla{\bf E}_{1}\bdot\mathbb{P}_{\rm gy} \;+\frac{}{} \nabla{\bf B}_{1}\bdot\mathbb{M}_{\rm gy}\right) .
   \label{eq:momentum_first_prime}
 \end{eqnarray}
In order to complete our proof, we now need to show that the right sides of Eqs.~\eqref{eq:momentum_first} and \eqref{eq:momentum_first_prime} are indeed equal to each other. For this purpose, we introduce the identities
\begin{eqnarray*}
\varrho_{\rm gy}\,{\bf E}_{1} + \frac{1}{c}\,{\bf J}_{\rm gy}\btimes{\bf B}_{1} & = &  \int_{\bf P} {\cal J}_{\rm gy}\,F \left( e\,\langle{\bf E}_{1{\rm gc}}\rangle \;+\; \frac{e}{c}\,\dot{\bf X}\btimes\langle{\bf B}_{1{\rm gc}}\rangle \right), \\
\nabla{\bf E}_{1}\bdot\mathbb{P}_{\rm gy} + \nabla{\bf B}_{1}\bdot\mathbb{M}_{\rm gy} & = & \int_{\bf P} {\cal J}_{\rm gy}\,F \left[ \left(\nabla{\bf P}_{1{\rm gy}} \;-\frac{}{} \nabla^{\prime}{\bf P}_{1{\rm gy}}\right)\bdot\dot{\bf X} \;-\; \left( \nabla K_{\rm gy} \;-\; \nabla^{\prime}K_{\rm gy} \right) \right],
\end{eqnarray*}
and by subtracting the right sides of Eqs.~\eqref{eq:momentum_first} and \eqref{eq:momentum_first_prime} from each other, we arrive at the identity
\begin{equation}
0 \;=\; \epsilon^{-1} \int_{\bf P} {\cal J}_{\rm gy}\,F \left( e\,{\bf E}_{\rm gy}^{*} \;+\; \frac{e}{c}\,\dot{\bf X}\btimes{\bf B}_{\rm gy}^{*} \;-\; \dot{p}_{\|}\;{\sf b}_{\rm gy}^{*} \right), 
\label{eq:mom_proof}
\end{equation}
which is identically satisfied as a result of the gyrocenter Euler-Lagrange equation \eqref{eq:EL_X}. We note that, just like the gyrokinetic energy conservation law, Eq.~\eqref{eq:mom_proof} is valid for any truncation order of the gyrocenter kinetic energy $K_{\rm gy}$. 

\subsubsection{Gyrokinetic angular-momentum conservation law}

Assuming now that the background magnetic field is axisymmetric, we derive the gyrokinetic canonical angular-momentum conservation law by taking the scalar product of Eq.~\eqref{eq:momentum} with $\partial{\bf x}/\partial\varphi$ (i.e., $\delta{\bf x} = \delta\varphi\;
\partial{\bf x}/\partial\varphi$), where the toroidal angle $\varphi$ is associated with rotations about the $z$-axis. Hence, the toroidal canonical 
angular-momentum density ${\cal P}_{{\rm gy}\varphi}^{*} \equiv  \vb{\cal P}_{{\rm gy}}^{*}\bdot\partial{\bf x}/\partial\varphi$ satisfies the Noether canonical angular-momentum equation
\begin{eqnarray}
\pd{{\cal P}_{{\rm gy}\varphi}^{*}}{t} + \nabla\vb{\cdot}\left({\sf T}_{{\rm gy}}^{*}\bdot\pd{\bf x}{\varphi}\right) &=& {\sf T}_{{\rm gy}}^{*\top}\;\vb{:}\;\nabla\left(\pd{\bf x}{\varphi}\right) - \pd{{\bf B}_{0}}{\varphi}\bdot\frac{\bf B}{4\pi} \nonumber \\
 &&+ \int_{\bf P} {\cal J}_{\rm gy}\,F\left( \frac{e}{c}\,\pd{{\bf A}_{0}^{*}}{\varphi}\bdot\dot{\bf X} + \epsilon\,\frac{\partial^{\prime}{\bf P}_{1{\rm gy}}}{\partial\varphi}\bdot\dot{\bf X} - \frac{\partial^{\prime}K_{\rm gy}}{\partial\varphi}\right),
 \label{eq:Pphi}
 \end{eqnarray}
where ${\sf T}_{{\rm gy}}^{*\top}$ denotes the transpose of the gyrokinetic stress tensor \eqref{eq:Tgy_can}. In addition, under the assumption that the background magnetic field is axisymmetric, we have $\partial B_{0}/\partial\varphi \equiv 0$ and we will use the identity $\partial\bhat_{0}/\partial\varphi \equiv \wh{\sf z}\btimes\bhat_{0}$, so that ${\bf B}\bdot\partial{\bf B}_{0}/\partial\varphi = \epsilon\,{\bf B}_{1}\bdot(\wh{\sf z}\btimes{\bf B}_{0})$. 

Instead of merely assuming that the right side of Eq.~\eqref{eq:Pphi} is zero, we will now systematically show how the various terms do cancel each other out. Before we begin, however, we note that the first term vanishes identically if the gyrokinetic stress tensor \eqref{eq:Tgy_can} is symmetric (i.e., ${\sf T}_{{\rm gy}}^{*\top} = {\sf T}_{{\rm gy}}^{*}$), which is expected (and required) when there is no separation between dynamical fields and background fields, e.g., in guiding-center Vlasov-Maxwell theory \citep{Brizard_Tronci_2016}.  

We now proceed with the proof that the right side of Eq.~\eqref{eq:Pphi} is zero. First, we note that since the dyadic tensor $\nabla(\partial{\bf x}/\partial\varphi) = \wh{R}\,\wh{\varphi} - \wh{\varphi}\,\wh{R}$ is  anti-symmetric (where $R \equiv
|\partial{\bf x}/\partial\varphi|$), only the anti-symmetric part of ${\sf T}_{{\rm gy}}^{*\top}$ contributes in the first term of Eq.~\eqref{eq:Pphi}:
\begin{eqnarray}
{\sf T}_{{\rm gy}}^{*\top}\;\vb{:}\;\nabla\left(\pd{\bf x}{\varphi}\right) & = & \wh{\sf z}\bdot\left[ \int_{\bf P} {\cal J}_{\rm gy}\,F\;\dot{\bf X}\btimes\left(\frac{e}{c}\,{\bf A}_{0}^{*} + \epsilon\,{\bf P}_{1{\rm gy}}\right) \;-\; \frac{\epsilon}{4\pi} \left(\mathbb{D}_{\rm gy}\btimes
{\bf E}_{1} \;+\frac{}{} {\bf B}_{1}\btimes\mathbb{H}_{\rm gy}\right) \right] \nonumber \\
 & = & \wh{\sf z}\bdot\left[ \int_{\bf P} {\cal J}_{\rm gy}\,F\;\dot{\bf X}\btimes\left(\frac{e}{c}\,{\bf A}_{0}^{*} + \epsilon\,{\bf P}_{1{\rm gy}}\right) \;+\; \epsilon\,{\bf E}_{1}\btimes\mathbb{P}_{\rm gy} \;+\; \epsilon\,{\bf B}_{1}\btimes\mathbb{M}_{\rm gy}\right] 
 \nonumber \\
  &&-\; \frac{\wh{\sf z}}{4\pi}\bdot(\epsilon\,{\bf B}_{1}\btimes{\bf B}_{0}),
\end{eqnarray} 
where we used the dyadic identity ${\bf V}{\bf W}:\nabla(\partial{\bf x}/\partial\varphi) \equiv \wh{\sf z}\bdot({\bf W}\btimes{\bf V})$, which holds for an arbitrary pair of vectors $({\bf V},{\bf W})$. Next, the last two terms are
\begin{eqnarray}
\frac{\partial^{\prime}{\bf P}_{1{\rm gy}}}{\partial\varphi}\vb{\cdot}\dot{\bf X} & = & \left[\left( \langle{\bf E}_{1{\rm gc}}\rangle + \frac{p_{\|}\bhat_{0}}{mc}\vb{\times}\langle{\bf B}_{1{\rm gc}}\rangle\right)\vb{\times}\pd{}{\varphi}\left(\frac{e\,\bhat_{0}}{\Omega_{0}}
\right) + \frac{p_{\|}}{mc}\left(\pd{\bhat_{0}}{\varphi}\vb{\times}\langle{\bf B}_{1{\rm gc}}\rangle\right)\vb{\times}\frac{e\,\bhat_{0}}{\Omega_{0}}\right]\bdot\dot{\bf X} \nonumber \\
 & = & \pd{}{\varphi}\left(\frac{e\,\bhat_{0}}{\Omega_{0}}\right) \bdot\left[\dot{\bf X}\btimes\left(\langle{\bf E}_{1{\rm gc}}\rangle \;+\; \frac{p_{\|}\bhat_{0}}{mc}\btimes\langle{\bf B}_{1{\rm gc}}\rangle\right) \right] \nonumber \\
  &&+\; \frac{p_{\|}}{mc}\left[\left(\wh{\sf z}\btimes\bhat_{0}\right)\btimes\langle{\bf B}_{1{\rm gc}}\rangle\right]\bdot\vb{\pi}_{\rm gy},
\end{eqnarray}
and
\begin{equation}
\frac{\partial^{\prime}K_{\rm gy}}{\partial\varphi} \;=\; \epsilon\,\mu\pd{\bhat_{0}}{\varphi}\bdot\langle\langle {\bf B}_{1{\rm gc}}\rangle\rangle \;+\; \epsilon^{2}\,\pd{\bhat_{0}}{\varphi}\bdot\left[\frac{p_{\|}{\bf B}_{1}}{B_{0}}\btimes
\frac{c}{B_{0}}\left({\bf E}_{1} \;+\; \frac{p_{\|}\bhat_{0}}{mc}\btimes{\bf B}_{1}\right) \right].
\end{equation}
Lastly, we write $\partial{\bf A}_{0}^{*}/\partial\varphi = \wh{\sf z}\btimes{\bf A}_{0}^{*}$ and, after some cancellations, Eq.~\eqref{eq:Pphi} becomes
\begin{eqnarray}
\pd{{\cal P}_{{\rm gy}\varphi}^{*}}{t} \;+\; \nabla\bdot\left({\sf T}_{{\rm gy}}^{*}\bdot\pd{\bf x}{\varphi}\right) & = & \epsilon\;\int_{\bf P} {\cal J}_{\rm gy}\,F\left[ \wh{\sf z}\bdot\left(\dot{\bf X}\btimes{\bf P}_{1{\rm gy}} \;-\; \mu\,\bhat_{0}\btimes
\langle\langle{\bf B}_{1{\rm gc}}\rangle\rangle\right) \right] \nonumber \\
 &&+\; \int_{\bf P} {\cal J}_{\rm gy}\,F \left( \epsilon\;\frac{\partial^{\prime}{\bf P}_{1{\rm gy}}}{\partial\varphi}\bdot\dot{\bf X}  \;-\; \epsilon^{2}\,\frac{\partial^{\prime}K_{2{\rm gy}}}{\partial\varphi} \right) \nonumber \\
  & &+\; \wh{\sf z}\bdot\epsilon\left({\bf E}_{1}\btimes\mathbb{P}_{\rm gy} \;+\frac{}{} {\bf B}_{1}\btimes\mathbb{M}_{\rm gy}\right),
 \label{eq:Pphi_2}
 \end{eqnarray}
where
\begin{eqnarray*}
\wh{\sf z}\bdot\epsilon\left({\bf E}_{1}\btimes\mathbb{P}_{\rm gy} \;+\frac{}{} {\bf B}_{1}\btimes\mathbb{M}_{\rm gy}\right) & = & \int_{\bf P} {\cal J}_{\rm gy}\,F\;\wh{\sf z}\bdot\epsilon\left( \langle{\bf E}_{1{\rm gc}}\rangle\btimes\vb{\pi}_{\rm gy} \;-\; \mu\,\langle\langle{\bf B}_{1{\rm gc}}\rangle\rangle\btimes\bhat_{0} \right) \\
 &&+\; \int_{\bf P} {\cal J}_{\rm gy}\,F\;\wh{\sf z}\bdot\epsilon\left[ \langle{\bf B}_{1{\rm gc}}\rangle\btimes\left(\vb{\pi}_{\rm gy}\btimes\frac{p_{\|}\bhat_{0}}{mc}\right) \right] \\
 &&-\; \epsilon^{2}\;\int_{\bf P} {\cal J}_{\rm gy}\,F\;\wh{\sf z}\bdot \left({\bf E}_{1}\btimes\pd{K_{2{\rm gy}}}{{\bf E}_{1}} \;+\; {\bf B}_{1}\btimes\pd{K_{2{\rm gy}}}{{\bf B}_{1}} \right),
 \end{eqnarray*}
with 
\begin{eqnarray*}
\wh{\sf z}\bdot \left({\bf E}_{1}\btimes\pd{K_{2{\rm gy}}}{{\bf E}_{1}} \;+\; {\bf B}_{1}\btimes\pd{K_{2{\rm gy}}}{{\bf B}_{1}} \right) &=& \wh{\sf z}\bdot \bhat_{0}\btimes\left[\frac{c}{B_{0}}\left({\bf E}_{1} \;+\; \frac{p_{\|}\bhat_{0}}{mc}\btimes{\bf B}_{1}\right) 
\btimes \frac{p_{\|}{\bf B}_{1}}{B_{0}}\right] \\
 &\equiv& -\;\frac{\partial^{\prime}K_{2{\rm gy}}}{\partial\varphi}.
 \end{eqnarray*}
We now write
\begin{eqnarray*}
\wh{\sf z}\bdot\left(\dot{\bf X}\vb{\times}{\bf P}_{1{\rm gy}}\right) &=& -\,\wh{\sf z}\bdot\left[ \left(\langle{\bf E}_{1{\rm gc}}\rangle \;+\; \frac{p_{\|}\bhat_{0}}{mc}\vb{\times}\langle{\bf B}_{1{\rm gc}}\rangle\right)\vb{\times}\vb{\pi}_{\rm gy} \right] \\
 &&-\; \pd{}{\varphi}\left(\frac{e\,\bhat_{0}}{\Omega_{0}}\right)\bdot\left[\dot{\bf X}\vb{\times}\left(\langle{\bf E}_{1{\rm gc}}\rangle \;+\; \frac{p_{\|}\bhat_{0}}{mc}\vb{\times}\langle{\bf B}_{1{\rm gc}}\rangle\right) \right],
 \end{eqnarray*}
so that, upon additional cancellations, Eq.~\eqref{eq:Pphi_2} becomes
\begin{eqnarray*} 
&& \pd{{\cal P}_{{\rm gy}\varphi}^{*}}{t} + \nabla\bdot\left({\sf T}_{{\rm gy}}^{*}\vb{\cdot}\pd{\bf x}{\varphi}\right) \\
&=& \epsilon\,\wh{\sf z}\vb{\cdot}\int_{\bf P} {\cal J}_{\rm gy}\,F\;\frac{p_{\|}}{mc}\left[ \langle{\bf B}_{1{\rm gc}}\rangle\vb{\times}\left(\vb{\pi}_{\rm gy}\vb{\times}\bhat_{0}\right) +
\vb{\pi}_{\rm gy}\vb{\times}\left(\bhat_{0}\vb{\times}\langle{\bf B}_{1{\rm gc}}\rangle\right) + \bhat_{0}\vb{\times}\left(\langle{\bf B}_{1{\rm gc}}\rangle\btimes\frac{}{}\vb{\pi}_{\rm gy}\right)\right].
\end{eqnarray*}
We finally obtain the gyrokinetic canonical angular-momentum conservation law
\begin{equation}
\pd{{\cal P}_{{\rm gy}\varphi}^{*}}{t} + \nabla\bdot\left({\sf T}_{{\rm gy}}^{*}\bdot\pd{\bf x}{\varphi}\right) \;=\; 0,
\label{eq:Pphi_final}
\end{equation}
upon using the Jacobi identity ${\bf A}\vb{\times}({\bf B}\vb{\times}{\bf C}) + {\bf B}\vb{\times}({\bf C}\vb{\times}{\bf A}) + {\bf C}\vb{\times}({\bf A}\vb{\times}{\bf B}) \equiv 0$  for the double vector product of any three arbitrary vector fields $({\bf A}, {\bf B}, 
{\bf C})$. We note that the terms involving the second-order gyrocenter kinetic energy $K_{2{\rm gy}}$ cancel each other out in the proof of the gyrokinetic canonical angular-momentum conservation law \eqref{eq:Pphi_final}, which implies that, like the conservation of gyrokinetic energy, the conservation of gyrokinetic canonical angular-momentum is valid if the gyrocenter kinetic energy is truncated at the first order.

In Eq.~\eqref{eq:Pphi_final}, the total toroidal angular-momentum density
\begin{eqnarray} 
{\cal P}^{*}_{{\rm gy}\varphi} & = & \int_{\bf P} {\cal J}_{\rm gy}\,F\,\left[ P_{{\rm gc}\varphi}^{*} \;+\; \epsilon\,\left(\langle{\bf E}_{1{\rm gc}}\rangle\btimes\frac{e\bhat_{0}}{\Omega_{0}}\bdot
\pd{\bf X}{\varphi} \;+\; \frac{p_{\|}}{B_{0}}\;\langle{\bf B}_{1\bot{\rm gc}}\rangle\bdot\pd{\bf X}{\varphi} \right)\right] \nonumber \\
 &&+\; \frac{\mathbb{D}_{\rm gy}}{4\pi c}\btimes\epsilon\,{\bf B}_{1}\bdot\pd{\bf X}{\varphi}
\label{eq:Pgy_can_phi} 
\end{eqnarray}
is the sum of the gyrocenter moment of the guiding-center toroidal angular-momentum $P_{{\rm gc}\varphi}^{*} = -\,(e/c)\,\psi + p_{\|}\,b_{0\varphi} + \cdots$, which is defined with higher-order guiding-center corrections as \citep{Tronko_Brizard_2015}
\begin{equation}
P_{{\rm gc}\varphi}^{*} \;\equiv\; \frac{e}{c}\,{\bf A}_{0}^{*}\bdot\pd{\bf X}{\varphi} \;=\; -\,\frac{e}{c} \left[ \psi \;+\; \nabla\bdot\left(\frac{J}{2\,m\,\Omega_{0}}\;\nabla\psi\right)\right] \;+\; p_{\|}\;b_{0\varphi} \;-\; 2\,J\;b_{0z},
\end{equation}
the toroidal components of the perturbed $E\times B$ velocity and magnetic-flutter momentum, and the toroidal component of the Minkowski electromagnetic momentum (which includes gyrocenter polarization effects). In the absence of magnetic-field perturbations, we recover the gyrokinetic toroidal angular-momentum density previously derived (without guiding-center corrections) in the electrostatic case \citep{TSH_2007,Scott_Smirnov_2010, Brizard_Tronko_2011}.

\section{\label{sec:Sum}Summary}

A new set of gyrokinetic Vlasov-Maxwell equations was derived according to a symplectic representation in which polarization effects were inserted in the symplectic structure. This new symplectic representation allowed for the introduction of self-consistent gyrocenter polarization and magnetization in the gyrokinetic Maxwell equations \eqref{eq:div_D}-\eqref{eq:curl_H} that contained contributions from the first-order gyrocenter symplectic structure as well as the gyrocenter Hamiltonian. The self-consistency of the gyrokinetic Vlasov-Maxwell equations \eqref{eq:Vlasov_gy}-\eqref{eq:curl_H} is guaranteed by their variational derivation from the gyrokinetic action functional \eqref{eq:A_gy}. By applying the Noether method on this gyrokinetic action functional, we were able to derive exact conservation laws for gyrokinetic energy as well as gyrokinetic toroidal angular-momentum under the assumption of a time-independent and axisymmetric background magnetic field. 

The numerical implementation of the gyrokinetic Vlasov-Maxwell equations \eqref{eq:Vlasov_gy}-\eqref{eq:curl_H} remains to be explored and is well outside of the scope of the present paper. In recent work \citep{Brizard_2019}, a truncated set of symplectic gyrokinetic equations was presented in which the second-order gyrocenter Hamiltonian \eqref{eq:H2_gy_ZLR} is omitted from the gyrocenter Hamiltonian. As was discussed on several occasions in Sec.~\ref{sec:symp_laws}, the gyrokinetic conservation laws
of energy-momentum and angular momentum remain exact when the second-order gyrocenter Hamiltonian \eqref{eq:H2_gy_ZLR} is omitted from the gyrokinetic Vlasov-Maxwell equations \eqref{eq:Vlasov_gy}-\eqref{eq:curl_H}. The numerical implementation of these truncated gyrokinetic Vlasov-Maxwell equations will also be explored in future work.

\acknowledgments

Part of the work presented here was carried out as part of a collaboration with the {\sf ELMFIRE} numerical simulation group at Aalto University (Finland). The Author acknowledges support from the National Science Foundation under contract No.~PHY-1805164.

\appendix

\section{\label{sec:Bessel}Bessel-function Identities}

In this Appendix, we use Bessel-function identities \citep{NIST_Bessel} to derive Eq.~\eqref{eq:Bessel_gy} under the assumption that the background magnetic field is uniform. We begin with the operator identity 
\begin{equation}
\langle {\sf T}_{\rm gc}^{-1}\rangle(z) \;\equiv\; \langle\exp(\vb{\rho}_{0}\bdot\nabla_{\bot})\rangle = J_{0}(z), 
\label{eq:J0}
\end{equation}
where $J_{\ell}(z)$ denotes the $\ell$th-order Bessel function with the argument $z$ defined from the relation 
\begin{equation}
z^{2} \;\equiv\; -\,(\,2J/m\Omega_{0})|\nabla_{\bot}|^{2}. 
\label{eq:z2_def}
\end{equation}
We note that the more conventional eikonal notation $\nabla_{\bot} = i\,{\bf k}_{\bot}$, for which $z^{2} = 2J|{\bf k}_{\bot}|^{2}/m\Omega_{0}$, is not necessary in what follows since derivatives with respect to $\nabla_{\bot}$ can be easily evaluated without ambiguity.

First, we derive an expression for $\langle\vb{\rho}_{0}\;{\sf T}_{\rm gc}^{-1}\rangle$:
\begin{equation}
\left\langle\vb{\rho}_{0}\frac{}{}{\sf T}_{\rm gc}^{-1}\right\rangle \;=\; \pd{\langle {\sf T}_{\rm gc}^{-1}\rangle}{\nabla_{\bot}} \;=\; \pd{z}{\nabla_{\bot}}\;J^{\prime}_{0}(z) \;=\; \left(-\,\frac{2}{z}\;\frac{J\,\nabla_{\bot}}{m\Omega_{0}}\right)\;J^{\prime}_{0}(z)  \;=\; 
\frac{J}{m\Omega_{0}}\left(\frac{J_{1}(z)}{z/2}\right)\;\nabla_{\bot},
\label{eq:rho_T}
\end{equation}
where we used $J_{0}^{\prime}(z) = -\,J_{1}(z)$. We can thus express
\begin{eqnarray}
-\,\frac{e\Omega_{0}}{c}\left\langle \pd{\vb{\rho}_{0}}{\zeta}\bdot{\bf A}_{1\bot{\rm gc}}\right\rangle &=& \frac{m\Omega_{0}^{2}}{B_{0}} \bhat_{0}\bdot\left\langle\vb{\rho}_{0}\,{\sf T}_{\rm gc}^{-1}\right\rangle\btimes{\bf A}_{1\bot} \;=\; \mu\;\left(\frac{J_{1}(z)}{z/2}\right)\,\bhat_{0}\bdot\nabla_{\bot}\btimes{\bf A}_{1\bot} \nonumber \\
 &\equiv& \mu\;\langle\langle B_{1\|{\rm gc}}\rangle\rangle,
 \end{eqnarray}
 where the symbol $\langle\langle \cdots\rangle\rangle$ is introduced \cite{Porazik_Lin_2011} to denote a gyro-surface average.

From Eq.~\eqref{eq:rho_T}, we obtain
\begin{eqnarray}
\pd{}{J}\left\langle\vb{\rho}_{0}\frac{}{}{\sf T}_{\rm gc}^{-1}\right\rangle  & = & -\;\pd{z}{J}\;\left(z\frac{}{}J_{1}(z)\right)^{\prime}\;\frac{\nabla_{\bot}}{|\nabla_{\bot}|^{2}} \;=\; -\;\pd{z}{J}\;\left(z\frac{}{}J_{0}(z)\right)\;\frac{\nabla_{\bot}}{|\nabla_{\bot}|^{2}}\nonumber \\
 & = & -\;\pd{(z^{2}/2)}{J}\;\frac{}{}J_{0}(z)\;\frac{\nabla_{\bot}}{|\nabla_{\bot}|^{2}} \;=\; \frac{J_{0}(z)}{m\Omega_{0}}\nabla_{\bot} \equiv \frac{1}{m\Omega_{0}}\nabla_{\bot}\langle{\sf T}_{\rm gc}^{-1}\rangle,
  \label{eq:mu_rho_T}
\end{eqnarray}
where we used the Bessel relation $(z\,J_{1}(z))^{\prime} = z\,J_{0}(z)$. Using this relation, we also obtain the gyro-surface average identity
\begin{equation}
\langle\langle {\sf T}_{\rm gc}^{-1}\rangle\rangle(z) \;\equiv\; \frac{2}{z^{2}}\;\int_{0}^{z}\langle{\sf T}_{\rm gc}^{-1}\rangle(\lambda)\;\lambda\,d\lambda \;=\; \frac{2}{z^{2}}\;\int_{0}^{z}J_{0}(\lambda)\;\lambda\,d\lambda \;=\; \frac{J_{1}(z)}{z/2}.
\end{equation}

Next, we derive the expression for $\langle\vb{\rho}_{0}\vb{\rho}_{0}\;{\sf T}_{\rm gc}^{-1}\rangle$:
\begin{eqnarray}
\left\langle\vb{\rho}_{0}\vb{\rho}_{0}\frac{}{}{\sf T}_{\rm gc}^{-1}\right\rangle & = & -\,\pd{}{\nabla_{\bot}}\left[\left(z\frac{}{}J_{1}(z)\right)\;\frac{\nabla_{\bot}}{|\nabla_{\bot}|^{2}} \right]  \nonumber \\
 &=& -\,\left(z\frac{}{}J_{1}(z)\right)\frac{\mathbb{I}_{\bot}}{|\nabla_{\bot}|^{2}} + \left(z^{2}\frac{}{}J_{2}(z)\right)\frac{\nabla_{\bot}\nabla_{\bot}}{|\nabla_{\bot}|^{4}},
\label{eq:rhorho_T}
\end{eqnarray}
where $\mathbb{I}_{\bot} \equiv \mathbb{I} - \bhat_{0}\bhat_{0}$ and we used the recurrence relation 
\[ z^{2}\,J_{2}(z) \;=\; 2z\,J_{1}(z) - z^{2}\,J_{0}(z). \]
We now derive the gyroaction derivative to find
\begin{eqnarray}
\pd{}{J}\left\langle\vb{\rho}_{0}\vb{\rho}_{0}\frac{}{}{\sf T}_{\rm gc}^{-1}\right\rangle & = & -\,\pd{z}{J}\;\left(z\frac{}{}J_{1}(z)\right)^{\prime}\frac{\mathbb{I}_{\bot}}{|\nabla_{\bot}|^{2}} \;+\; \pd{z}{J}\;\left(z^{2}\frac{}{}J_{2}(z)\right)^{\prime}\frac{\nabla_{\bot}\nabla_{\bot}}{|\nabla_{\bot}|^{4}} \\
  & = & -\,\pd{(z^{2}/2)}{J}\;J_{0}(z)\;\frac{\mathbb{I}_{\bot}}{|\nabla_{\bot}|^{2}} \;+\; \pd{(z^{2}/2)}{J}\;z\,J_{1}(z)\;\frac{\nabla_{\bot}\nabla_{\bot}}{|\nabla_{\bot}|^{4}} \nonumber \\
  & = & \frac{J_{0}(z)}{m\Omega_{0}}\;\mathbb{I}_{\bot} \;+\; \frac{2\,J_{1}(z)}{z\,m\Omega_{0}} \left( \frac{J}{m\Omega_{0}}\,\nabla_{\bot}\nabla_{\bot}\right),
  \nonumber
\end{eqnarray}
where we used $(z^{2}\,J_{2}(z))^{\prime} = z^{2}\,J_{1}(z)$.

We now combine these results to obtain the formula found in Eq.~\eqref{eq:Bessel_gy}:
\begin{eqnarray}
\pd{}{J}\left\langle \vb{\rho}_{0}\,\pd{S_{1}}{\zeta}\right\rangle & = & \frac{e}{\Omega_{0}}\pd{}{J}\left\langle\vb{\rho}_{0}\,{\sf T}_{\rm gc}^{-1}\right\rangle \left( \Phi_{1} - \frac{v_{\|}}{c}\,A_{1\|} \right)  \;-\;\frac{e}{c}\pd{}{J}\left\langle\vb{\rho}_{0}\vb{\rho}_{0}\frac{}{}{\sf T}_{\rm gc}^{-1}\right\rangle\bdot\bhat_{0}\btimes{\bf A}_{1\bot} \nonumber \\
  & = & \frac{e}{m\Omega_{0}^{2}}\left(\nabla_{\bot}\langle\Phi_{1{\rm gc}}\rangle - \frac{v_{\|}}{c}\,\nabla_{\bot}\langle A_{1\|{\rm gc}}\rangle\right) \;-\; \frac{\bhat_{0}}{B_{0}}\btimes  \langle{\bf A}_{1\bot{\rm gc}}\rangle \nonumber \\
  &&+\; \frac{\mu}{m\Omega_{0}^{2}}\,\nabla_{\bot}\langle\langle B_{1\|{\rm gc}}\rangle\rangle.
   \label{eq:App_id}
\end{eqnarray}

\bibliography{symp}

\begin{thebibliography}{47}
\expandafter\ifx\csname natexlab\endcsname\relax\def\natexlab#1{#1}\fi
\expandafter\ifx\csname bibnamefont\endcsname\relax
  \def\bibnamefont#1{#1}\fi
\expandafter\ifx\csname bibfnamefont\endcsname\relax
  \def\bibfnamefont#1{#1}\fi
\expandafter\ifx\csname citenamefont\endcsname\relax
  \def\citenamefont#1{#1}\fi
\expandafter\ifx\csname url\endcsname\relax
  \def\url#1{\texttt{#1}}\fi
\expandafter\ifx\csname urlprefix\endcsname\relax\def\urlprefix{URL }\fi
\providecommand{\bibinfo}[2]{#2}
\providecommand{\eprint}[2][]{\url{#2}}

\bibitem[{\citenamefont{Taylor}(1967)}]{Taylor_1967}
\bibinfo{author}{\bibfnamefont{J.~B.} \bibnamefont{Taylor}},
  \bibinfo{journal}{Phys. Fluids} \textbf{\bibinfo{volume}{10}},
  \bibinfo{pages}{1357} (\bibinfo{year}{1967}).

\bibitem[{\citenamefont{Cary and Brizard}(2009)}]{Cary_Brizard_2009}
\bibinfo{author}{\bibfnamefont{J.~R.} \bibnamefont{Cary}} \bibnamefont{and}
  \bibinfo{author}{\bibfnamefont{A.~J.} \bibnamefont{Brizard}},
  \bibinfo{journal}{Rev. Mod. Phys.} \textbf{\bibinfo{volume}{81}},
  \bibinfo{pages}{693} (\bibinfo{year}{2009}).

\bibitem[{\citenamefont{Catto}(1978)}]{Catto_1978}
\bibinfo{author}{\bibfnamefont{P.~J.} \bibnamefont{Catto}},
  \bibinfo{journal}{Plasma Phys.} \textbf{\bibinfo{volume}{20}},
  \bibinfo{pages}{719} (\bibinfo{year}{1978}).

\bibitem[{\citenamefont{Catto et~al.}(1981)\citenamefont{Catto, Tang, and
  Baldwin}}]{Catto_1981}
\bibinfo{author}{\bibfnamefont{P.~J.} \bibnamefont{Catto}},
  \bibinfo{author}{\bibfnamefont{W.~M.} \bibnamefont{Tang}}, \bibnamefont{and}
  \bibinfo{author}{\bibfnamefont{D.~E.} \bibnamefont{Baldwin}},
  \bibinfo{journal}{Plasma Phys.} \textbf{\bibinfo{volume}{23}},
  \bibinfo{pages}{639} (\bibinfo{year}{1981}).

\bibitem[{\citenamefont{Frieman and Chen}(1982)}]{Frieman_Chen_1982}
\bibinfo{author}{\bibfnamefont{E.~A.} \bibnamefont{Frieman}} \bibnamefont{and}
  \bibinfo{author}{\bibfnamefont{L.}~\bibnamefont{Chen}},
  \bibinfo{journal}{Phys. Fluids} \textbf{\bibinfo{volume}{25}},
  \bibinfo{pages}{502} (\bibinfo{year}{1982}).

\bibitem[{\citenamefont{Brizard and Hahm}(2007)}]{Brizard_Hahm_2007}
\bibinfo{author}{\bibfnamefont{A.~J.} \bibnamefont{Brizard}} \bibnamefont{and}
  \bibinfo{author}{\bibfnamefont{T.~S.} \bibnamefont{Hahm}},
  \bibinfo{journal}{Rev. Mod. Phys.} \textbf{\bibinfo{volume}{79}},
  \bibinfo{pages}{421} (\bibinfo{year}{2007}).

\bibitem[{\citenamefont{Garbet et~al.}(2010)\citenamefont{Garbet, Idomura,
  Villard, and Watanabe}}]{Garbet_2010}
\bibinfo{author}{\bibfnamefont{X.}~\bibnamefont{Garbet}},
  \bibinfo{author}{\bibfnamefont{Y.}~\bibnamefont{Idomura}},
  \bibinfo{author}{\bibfnamefont{L.}~\bibnamefont{Villard}}, \bibnamefont{and}
  \bibinfo{author}{\bibfnamefont{T.~H.} \bibnamefont{Watanabe}},
  \bibinfo{journal}{Nuc. Fusion} \textbf{\bibinfo{volume}{50}},
  \bibinfo{pages}{043002} (\bibinfo{year}{2010}).

\bibitem[{\citenamefont{Krommes}(2012)}]{Krommes_2012}
\bibinfo{author}{\bibfnamefont{J.~A.} \bibnamefont{Krommes}},
  \bibinfo{journal}{Annu. Rev. Fluid Mech} \textbf{\bibinfo{volume}{44}},
  \bibinfo{pages}{175} (\bibinfo{year}{2012}).

\bibitem[{\citenamefont{Dubin et~al.}(1983)\citenamefont{Dubin, Krommes,
  Oberman, and Lee}}]{Dubin_1983}
\bibinfo{author}{\bibfnamefont{D.~H.~E.} \bibnamefont{Dubin}},
  \bibinfo{author}{\bibfnamefont{J.~A.} \bibnamefont{Krommes}},
  \bibinfo{author}{\bibfnamefont{C.}~\bibnamefont{Oberman}}, \bibnamefont{and}
  \bibinfo{author}{\bibfnamefont{W.~W.} \bibnamefont{Lee}},
  \bibinfo{journal}{Phys. Fluids} \textbf{\bibinfo{volume}{26}},
  \bibinfo{pages}{3524} (\bibinfo{year}{1983}).

\bibitem[{\citenamefont{Hahm et~al.}(1988)\citenamefont{Hahm, Lee, and
  Brizard}}]{HLB_1988}
\bibinfo{author}{\bibfnamefont{T.~S.} \bibnamefont{Hahm}},
  \bibinfo{author}{\bibfnamefont{W.~W.} \bibnamefont{Lee}}, \bibnamefont{and}
  \bibinfo{author}{\bibfnamefont{A.~J.} \bibnamefont{Brizard}},
  \bibinfo{journal}{Phys. Fluids} \textbf{\bibinfo{volume}{31}},
  \bibinfo{pages}{1940} (\bibinfo{year}{1988}).

\bibitem[{\citenamefont{Brizard}(1989)}]{Brizard_1989}
\bibinfo{author}{\bibfnamefont{A.~J.} \bibnamefont{Brizard}},
  \bibinfo{journal}{J. Plasma Phys.} \textbf{\bibinfo{volume}{41}},
  \bibinfo{pages}{541} (\bibinfo{year}{1989}).

\bibitem[{\citenamefont{Sugama}(2000)}]{Sugama_2000}
\bibinfo{author}{\bibfnamefont{H.}~\bibnamefont{Sugama}},
  \bibinfo{journal}{Phys. Plasmas} \textbf{\bibinfo{volume}{7}},
  \bibinfo{pages}{466} (\bibinfo{year}{2000}).

\bibitem[{\citenamefont{Brizard}(2000{\natexlab{a}})}]{Brizard_PRL_2000}
\bibinfo{author}{\bibfnamefont{A.~J.} \bibnamefont{Brizard}},
  \bibinfo{journal}{Phys. Rev. Lett.} \textbf{\bibinfo{volume}{84}},
  \bibinfo{pages}{5768} (\bibinfo{year}{2000}{\natexlab{a}}).

\bibitem[{\citenamefont{Brizard}(2000{\natexlab{b}})}]{Brizard_PoP_2000}
\bibinfo{author}{\bibfnamefont{A.~J.} \bibnamefont{Brizard}},
  \bibinfo{journal}{Phys. Plasmas} \textbf{\bibinfo{volume}{7}},
  \bibinfo{pages}{4816} (\bibinfo{year}{2000}{\natexlab{b}}).

\bibitem[{\citenamefont{Brizard}(2010)}]{Brizard_2010}
\bibinfo{author}{\bibfnamefont{A.~J.} \bibnamefont{Brizard}},
  \bibinfo{journal}{Phys. Plasmas} \textbf{\bibinfo{volume}{17}},
  \bibinfo{pages}{042303} (\bibinfo{year}{2010}).

\bibitem[{\citenamefont{Mandell et~al.}(2020)\citenamefont{Mandell, Hakim,
  Hammett, and Francisquez}}]{Mandell_2020}
\bibinfo{author}{\bibfnamefont{N.~R.} \bibnamefont{Mandell}},
  \bibinfo{author}{\bibfnamefont{A.}~\bibnamefont{Hakim}},
  \bibinfo{author}{\bibfnamefont{G.~W.} \bibnamefont{Hammett}},
  \bibnamefont{and}
  \bibinfo{author}{\bibfnamefont{M.}~\bibnamefont{Francisquez}},
  \bibinfo{journal}{J. Plasma Phys.} \textbf{\bibinfo{volume}{86}},
  \bibinfo{pages}{905860109} (\bibinfo{year}{2020}).

\bibitem[{\citenamefont{Brizard}(2008)}]{Brizard_2008}
\bibinfo{author}{\bibfnamefont{A.~J.} \bibnamefont{Brizard}},
  \bibinfo{journal}{Comm. Nonlin. Sci. Num. Sim.}
  \textbf{\bibinfo{volume}{13}}, \bibinfo{pages}{24} (\bibinfo{year}{2008}).

\bibitem[{\citenamefont{Brizard}(2009)}]{Brizard_2009}
\bibinfo{author}{\bibfnamefont{A.~J.} \bibnamefont{Brizard}},
  \bibinfo{journal}{J. Phys. Conf. ser.} \textbf{\bibinfo{volume}{169}},
  \bibinfo{pages}{012003} (\bibinfo{year}{2009}).

\bibitem[{\citenamefont{Brizard and Tronci}(2016)}]{Brizard_Tronci_2016}
\bibinfo{author}{\bibfnamefont{A.~J.} \bibnamefont{Brizard}} \bibnamefont{and}
  \bibinfo{author}{\bibfnamefont{C.}~\bibnamefont{Tronci}},
  \bibinfo{journal}{Phys. Plasmas} \textbf{\bibinfo{volume}{23}},
  \bibinfo{pages}{062107} (\bibinfo{year}{2016}).

\bibitem[{\citenamefont{Brizard}(2018)}]{Brizard_2018}
\bibinfo{author}{\bibfnamefont{A.~J.} \bibnamefont{Brizard}},
  \bibinfo{journal}{Phys. Plasmas} \textbf{\bibinfo{volume}{25}},
  \bibinfo{pages}{112112} (\bibinfo{year}{2018}).

\bibitem[{\citenamefont{Hirvijoki et~al.}(2020)\citenamefont{Hirvijoki, Burby,
  Pfefferl\'{e}, and Brizard}}]{EH_2020}
\bibinfo{author}{\bibfnamefont{E.}~\bibnamefont{Hirvijoki}},
  \bibinfo{author}{\bibfnamefont{J.}~\bibnamefont{Burby}},
  \bibinfo{author}{\bibfnamefont{D.}~\bibnamefont{Pfefferl\'{e}}},
  \bibnamefont{and} \bibinfo{author}{\bibfnamefont{A.~J.}
  \bibnamefont{Brizard}}, \bibinfo{journal}{J. Phys. A: Theoretical and
  Mathematical} \textbf{\bibinfo{volume}{53}}, \bibinfo{pages}{235204}
  (\bibinfo{year}{2020}).

\bibitem[{\citenamefont{Pfirsch}(1984)}]{Pfirsch_1984}
\bibinfo{author}{\bibfnamefont{D.}~\bibnamefont{Pfirsch}}, \bibinfo{journal}{Z.
  Naturforsch. a} \textbf{\bibinfo{volume}{39}}, \bibinfo{pages}{1}
  (\bibinfo{year}{1984}).

\bibitem[{\citenamefont{Pfirsch and Morrison}(1985)}]{Pfirsch_Morrison_1985}
\bibinfo{author}{\bibfnamefont{D.}~\bibnamefont{Pfirsch}} \bibnamefont{and}
  \bibinfo{author}{\bibfnamefont{P.~J.} \bibnamefont{Morrison}},
  \bibinfo{journal}{Phys. Rev. A} \textbf{\bibinfo{volume}{32}},
  \bibinfo{pages}{1714} (\bibinfo{year}{1985}).

\bibitem[{\citenamefont{Dimits et~al.}(1992)\citenamefont{Dimits, LoDestro, and
  Dubin}}]{Dimits_1992}
\bibinfo{author}{\bibfnamefont{A.~M.} \bibnamefont{Dimits}},
  \bibinfo{author}{\bibfnamefont{L.~L.} \bibnamefont{LoDestro}},
  \bibnamefont{and} \bibinfo{author}{\bibfnamefont{D.~H.~E.}
  \bibnamefont{Dubin}}, \bibinfo{journal}{Phys. Fluids B}
  \textbf{\bibinfo{volume}{4}}, \bibinfo{pages}{274} (\bibinfo{year}{1992}).

\bibitem[{\citenamefont{Brizard}(2013)}]{Brizard_2013}
\bibinfo{author}{\bibfnamefont{A.~J.} \bibnamefont{Brizard}},
  \bibinfo{journal}{Phys. Plasmas} \textbf{\bibinfo{volume}{20}},
  \bibinfo{pages}{092309} (\bibinfo{year}{2013}).

\bibitem[{\citenamefont{Tronko and Brizard}(2015)}]{Tronko_Brizard_2015}
\bibinfo{author}{\bibfnamefont{N.}~\bibnamefont{Tronko}} \bibnamefont{and}
  \bibinfo{author}{\bibfnamefont{A.~J.} \bibnamefont{Brizard}},
  \bibinfo{journal}{Phys. Plasmas} \textbf{\bibinfo{volume}{22}},
  \bibinfo{pages}{112507} (\bibinfo{year}{2015}).

\bibitem[{\citenamefont{Brizard}(2017{\natexlab{a}})}]{Brizard_2017a}
\bibinfo{author}{\bibfnamefont{A.~J.} \bibnamefont{Brizard}},
  \bibinfo{journal}{Phys. Plasmas} \textbf{\bibinfo{volume}{24}},
  \bibinfo{pages}{042115} (\bibinfo{year}{2017}{\natexlab{a}}).

\bibitem[{\citenamefont{Littlejohn}(1983)}]{Littlejohn_1983}
\bibinfo{author}{\bibfnamefont{R.~G.} \bibnamefont{Littlejohn}},
  \bibinfo{journal}{J. Plasma Phys.} \textbf{\bibinfo{volume}{29}},
  \bibinfo{pages}{111} (\bibinfo{year}{1983}).

\bibitem[{\citenamefont{Pfirsch and Correa-Restrepo}(2004)}]{PCR_2004}
\bibinfo{author}{\bibfnamefont{D.}~\bibnamefont{Pfirsch}} \bibnamefont{and}
  \bibinfo{author}{\bibfnamefont{D.}~\bibnamefont{Correa-Restrepo}},
  \bibinfo{journal}{J. Plasma Phys.} \textbf{\bibinfo{volume}{70}},
  \bibinfo{pages}{719} (\bibinfo{year}{2004}).

\bibitem[{\citenamefont{Brizard}(2017{\natexlab{b}})}]{Brizard_2017b}
\bibinfo{author}{\bibfnamefont{A.~J.} \bibnamefont{Brizard}},
  \bibinfo{journal}{Phys. Plasmas} \textbf{\bibinfo{volume}{24}},
  \bibinfo{pages}{081201} (\bibinfo{year}{2017}{\natexlab{b}}).

\bibitem[{\citenamefont{Wang and Hahm}(2010{\natexlab{a}})}]{Wang_Hahm_2010a}
\bibinfo{author}{\bibfnamefont{L.}~\bibnamefont{Wang}} \bibnamefont{and}
  \bibinfo{author}{\bibfnamefont{T.~S.} \bibnamefont{Hahm}},
  \bibinfo{journal}{Phys. Plasmas} \textbf{\bibinfo{volume}{17}},
  \bibinfo{pages}{082304} (\bibinfo{year}{2010}{\natexlab{a}}).

\bibitem[{\citenamefont{Wang and Hahm}(2010{\natexlab{b}})}]{Wang_Hahm_2010b}
\bibinfo{author}{\bibfnamefont{L.}~\bibnamefont{Wang}} \bibnamefont{and}
  \bibinfo{author}{\bibfnamefont{T.~S.} \bibnamefont{Hahm}},
  \bibinfo{journal}{Phys. Plasmas} \textbf{\bibinfo{volume}{17}},
  \bibinfo{pages}{124702} (\bibinfo{year}{2010}{\natexlab{b}}).

\bibitem[{\citenamefont{Leering et~al.}(2010)\citenamefont{Leering, Parra, and
  Heikkinen}}]{Leerink_2010}
\bibinfo{author}{\bibfnamefont{S.}~\bibnamefont{Leering}},
  \bibinfo{author}{\bibfnamefont{F.~I.} \bibnamefont{Parra}}, \bibnamefont{and}
  \bibinfo{author}{\bibfnamefont{J.~A.} \bibnamefont{Heikkinen}},
  \bibinfo{journal}{Phys. Plasmas} \textbf{\bibinfo{volume}{17}},
  \bibinfo{pages}{124701} (\bibinfo{year}{2010}).

\bibitem[{\citenamefont{Heikkinen and Nora}(2011)}]{Heikkinen_Nora_2011}
\bibinfo{author}{\bibfnamefont{J.~A.} \bibnamefont{Heikkinen}}
  \bibnamefont{and} \bibinfo{author}{\bibfnamefont{M.}~\bibnamefont{Nora}},
  \bibinfo{journal}{Phys. Plasmas} \textbf{\bibinfo{volume}{18}},
  \bibinfo{pages}{022310} (\bibinfo{year}{2011}).

\bibitem[{\citenamefont{Duthoit et~al.}(2014)\citenamefont{Duthoit, Hahm, and
  Wang}}]{Duthoit_2014}
\bibinfo{author}{\bibfnamefont{F.-X.} \bibnamefont{Duthoit}},
  \bibinfo{author}{\bibfnamefont{T.~S.} \bibnamefont{Hahm}}, \bibnamefont{and}
  \bibinfo{author}{\bibfnamefont{L.}~\bibnamefont{Wang}},
  \bibinfo{journal}{Phys. Plasmas} \textbf{\bibinfo{volume}{21}},
  \bibinfo{pages}{082301} (\bibinfo{year}{2014}).

\bibitem[{\citenamefont{Burby and Brizard}(2019)}]{Burby_Brizard_2019}
\bibinfo{author}{\bibfnamefont{J.~W.} \bibnamefont{Burby}} \bibnamefont{and}
  \bibinfo{author}{\bibfnamefont{A.~J.} \bibnamefont{Brizard}},
  \bibinfo{journal}{Phys. Lett. A} \textbf{\bibinfo{volume}{383}},
  \bibinfo{pages}{2172} (\bibinfo{year}{2019}).

\bibitem[{\citenamefont{Littlejohn}(1982)}]{Littlejohn_1982}
\bibinfo{author}{\bibfnamefont{R.~G.} \bibnamefont{Littlejohn}},
  \bibinfo{journal}{J. Math. Phys.} \textbf{\bibinfo{volume}{23}},
  \bibinfo{pages}{742} (\bibinfo{year}{1982}).

\bibitem[{\citenamefont{Brizard et~al.}(2016)\citenamefont{Brizard, Morrison,
  Burby, de~Guillebon, and Vittot}}]{AJB_2016}
\bibinfo{author}{\bibfnamefont{A.~J.} \bibnamefont{Brizard}},
  \bibinfo{author}{\bibfnamefont{P.~J.} \bibnamefont{Morrison}},
  \bibinfo{author}{\bibfnamefont{J.~W.} \bibnamefont{Burby}},
  \bibinfo{author}{\bibfnamefont{L.}~\bibnamefont{de~Guillebon}},
  \bibnamefont{and} \bibinfo{author}{\bibfnamefont{M.}~\bibnamefont{Vittot}},
  \bibinfo{journal}{J. Plasma Phys.} \textbf{\bibinfo{volume}{82}},
  \bibinfo{pages}{905820608} (\bibinfo{year}{2016}).

\bibitem[{\citenamefont{Porazik and Lin}(2011)}]{Porazik_Lin_2011}
\bibinfo{author}{\bibfnamefont{P.}~\bibnamefont{Porazik}} \bibnamefont{and}
  \bibinfo{author}{\bibfnamefont{Z.}~\bibnamefont{Lin}},
  \bibinfo{journal}{Commun. Comput. Phys.} \textbf{\bibinfo{volume}{10}},
  \bibinfo{pages}{899} (\bibinfo{year}{2011}).

\bibitem[{\citenamefont{Correa-Restrepo and Pfirsch}(2004)}]{CRP_2004}
\bibinfo{author}{\bibfnamefont{D.}~\bibnamefont{Correa-Restrepo}}
  \bibnamefont{and} \bibinfo{author}{\bibfnamefont{D.}~\bibnamefont{Pfirsch}},
  \bibinfo{journal}{J. Plasma Phys.} \textbf{\bibinfo{volume}{70}},
  \bibinfo{pages}{757} (\bibinfo{year}{2004}).

\bibitem[{\citenamefont{Squire et~al.}(2013)\citenamefont{Squire, Qin, Tang,
  and Chandre}}]{Squire_2013}
\bibinfo{author}{\bibfnamefont{J.}~\bibnamefont{Squire}},
  \bibinfo{author}{\bibfnamefont{H.}~\bibnamefont{Qin}},
  \bibinfo{author}{\bibfnamefont{W.~M.} \bibnamefont{Tang}}, \bibnamefont{and}
  \bibinfo{author}{\bibfnamefont{C.}~\bibnamefont{Chandre}},
  \bibinfo{journal}{Phys. Plasmas} \textbf{\bibinfo{volume}{20}},
  \bibinfo{pages}{022501} (\bibinfo{year}{2013}).

\bibitem[{\citenamefont{Scott and Smirnov}(2010)}]{Scott_Smirnov_2010}
\bibinfo{author}{\bibfnamefont{B.}~\bibnamefont{Scott}} \bibnamefont{and}
  \bibinfo{author}{\bibfnamefont{J.}~\bibnamefont{Smirnov}},
  \bibinfo{journal}{Phys. Plasmas} \textbf{\bibinfo{volume}{17}},
  \bibinfo{pages}{112302} (\bibinfo{year}{2010}).

\bibitem[{\citenamefont{Brizard and Tronko}(2011)}]{Brizard_Tronko_2011}
\bibinfo{author}{\bibfnamefont{A.~J.} \bibnamefont{Brizard}} \bibnamefont{and}
  \bibinfo{author}{\bibfnamefont{N.}~\bibnamefont{Tronko}},
  \bibinfo{journal}{Phys. Plasmas} \textbf{\bibinfo{volume}{18}},
  \bibinfo{pages}{082307} (\bibinfo{year}{2011}).

\bibitem[{\citenamefont{Burby et~al.}(2015)\citenamefont{Burby, Brizard,
  Morrison, and Qin}}]{Burby_2015}
\bibinfo{author}{\bibfnamefont{J.~W.} \bibnamefont{Burby}},
  \bibinfo{author}{\bibfnamefont{A.~J.} \bibnamefont{Brizard}},
  \bibinfo{author}{\bibfnamefont{P.~J.} \bibnamefont{Morrison}},
  \bibnamefont{and} \bibinfo{author}{\bibfnamefont{H.}~\bibnamefont{Qin}},
  \bibinfo{journal}{Phys. Lett. A} \textbf{\bibinfo{volume}{379}},
  \bibinfo{pages}{2073} (\bibinfo{year}{2015}).

\bibitem[{\citenamefont{Hahm et~al.}(2007)\citenamefont{Hahm, Diamond, Gurcan,
  and Rewoldt}}]{TSH_2007}
\bibinfo{author}{\bibfnamefont{T.~S.} \bibnamefont{Hahm}},
  \bibinfo{author}{\bibfnamefont{P.~H.} \bibnamefont{Diamond}},
  \bibinfo{author}{\bibfnamefont{O.~D.} \bibnamefont{Gurcan}},
  \bibnamefont{and} \bibinfo{author}{\bibfnamefont{G.}~\bibnamefont{Rewoldt}},
  \bibinfo{journal}{Phys. Plasmas} \textbf{\bibinfo{volume}{14}},
  \bibinfo{pages}{072302} (\bibinfo{year}{2007}).

\bibitem[{\citenamefont{Brizard}(2019)}]{Brizard_2019}
\bibinfo{author}{\bibfnamefont{A.~J.} \bibnamefont{Brizard}},
  \bibinfo{journal}{arXiv:1907.11204}  (\bibinfo{year}{2019}).

\bibitem[{\citenamefont{Olver and Maximon}(2010)}]{NIST_Bessel}
\bibinfo{author}{\bibfnamefont{F.~W.~J.} \bibnamefont{Olver}} \bibnamefont{and}
  \bibinfo{author}{\bibfnamefont{L.~C.} \bibnamefont{Maximon}}, in
  \emph{\bibinfo{booktitle}{NIST {H}andbook of {M}athematical {F}unctions}}
  (\bibinfo{publisher}{Cambridge University Press}, \bibinfo{year}{2010}).

\end{thebibliography}

\end{document}